\documentclass{PoS}

\usepackage{amsmath}
\usepackage{color}
\let\normalcolor\relax

\newcommand{\be}{\begin{equation}}
\newcommand{\ee}{\end{equation}}
\newcommand{\benn}{\nonumber\begin{equation}}
\newcommand{\eenn}{\nonumber\end{equation}}
\def\bea{\begin{eqnarray}} \def\eea{\end{eqnarray}}
\def\beann{\begin{eqnarray*}} \def\eeann{\end{eqnarray*}}

\def\lsim{\raise0.3ex\hbox{$<$\kern-0.75em\raise-1.1ex\hbox{$\sim$}}}
\def\gsim{\raise0.3ex\hbox{$>$\kern-0.75em\raise-1.1ex\hbox{$\sim$}}}
\def\slash#1{#1\!\!\!\!\!/\!\,\,}
\def\Dslash{\slash D}

\title{\vspace*{-1cm}
\begin{flushright}
\texttt{\footnotesize CERN-PH-TH/2010-090}
\end{flushright}
\vfill
Simulating QCD at finite density
}

\ShortTitle{Finite mu}

\author{\speaker{Philippe de Forcrand} \\ 
        Institute for Theoretical Physics, ETH Z\"urich, CH-8093 Z\"urich, Switzerland \\
        and \\
        CERN, Physics Department, TH Unit, CH-1211 Geneva 23, Switzerland \\
        E-mail: \email{forcrand@phys.ethz.ch}}

\abstract{
In this review, I recall the nature and the inevitability of the ``sign problem'' which plagues
attempts to simulate lattice QCD at finite baryon density. I present the main approaches used
to circumvent the sign problem at small chemical potential. I sketch how 
one can predict analytically the severity of the sign problem, as well as the numerically accessible range of 
baryon densities. I review progress towards the determination of the pseudo-critical temperature
$T_c(\mu)$, and towards the identification of a possible QCD critical point.
Some promising advances with non-standard approaches are reviewed.
}

\FullConference{The XXVII International Symposium on Lattice Field Theory - LAT2009\\
		 July 26-31 2009\\
		 Peking University, Beijing, China}

\begin{document}


\section{Introduction}

\begin{figure}
\centerline{
\includegraphics[width=0.80\textwidth]{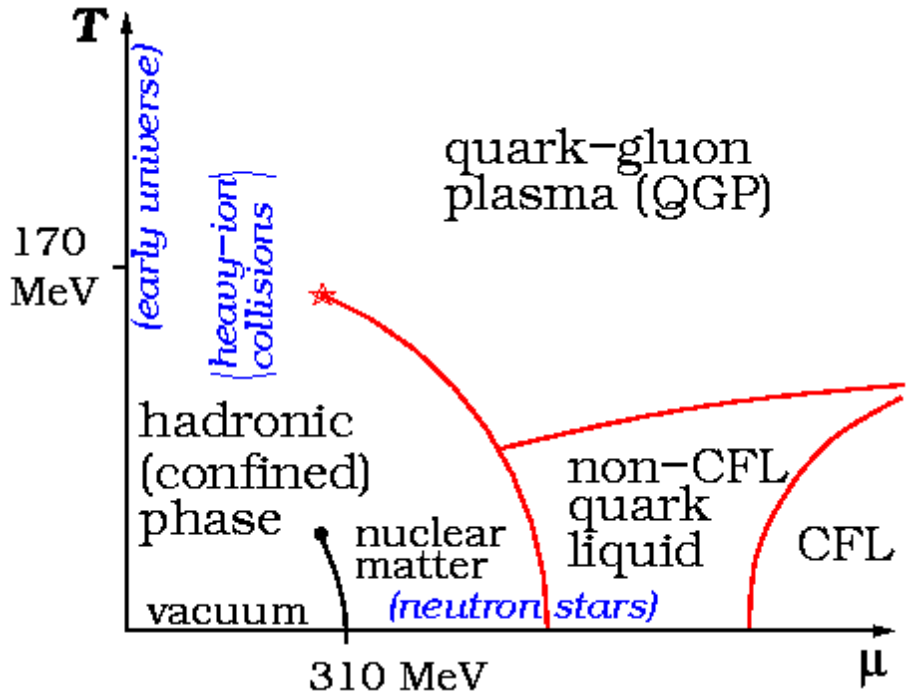}
}
\caption{Conjectured phase diagram of QCD as a function of quark chemical potential $\mu$ and temperature $T$, from Wikipedia.}
\label{fig:phasediag}
\end{figure}

Just like with water, the form taken by quark matter depends on its temperature and its density,
or equivalently the chemical potential coupled to the quark number.
In fact, one should consider distinct, possibly different chemical potentials coupled to the densities of
$u, d,$ and $s$ quarks which are separately conserved by the strong interactions, 
giving in total a 4-dimensional parameter space with a rich phase diagram.
A conjectured two-dimensional $(\mu,T)$ section (where the requirements of electric neutrality and of beta-equilibrium reduce the
three chemical potentials to a single combination) is proposed Fig.~\ref{fig:phasediag},
taken from a popular reference. The behaviour of QCD in some limiting cases (high $T$ or large $\mu$)
can be predicted from perturbation theory thanks to asymptotic freedom, and the $T\geq 0, \mu=0$ 
properties have been well studied on the lattice. 
Otherwise, almost all of the phase diagram Fig.~\ref{fig:phasediag} is based on
educated guesses awaiting validation. Putting this phase diagram on a firm basis is obviously of fundamental
importance. 

The QCD phase diagram follows from the non-perturbative properties of the QCD Lagrangian, and can in
principle be determined unambiguously by lattice simulations. Unfortunately, standard Monte Carlo 
simulations can only be applied to the $\mu=0$ vertical axis in Fig.~\ref{fig:phasediag}. As is well-known, for $\mu\neq 0$
the simulations are plagued by the ``sign problem''. The purpose of this review is to explain the origin
and the nature of the sign problem, and to report on recent progress towards circumventing it.
I have tried both to start from elementary considerations and to cover some very recent, promising developments.
This has forced me to skip reviewing some new work presented at this Conference, for which I apologize.

\section{Sign problem}
\label{sec:sign}

The sign problem is a necessary evil, unavoidable as soon as one integrates out the fermion fields
and expresses the partition function in terms of the gauge fields. 
Analytic integration over each fermion species gives a factor $\det(\Dslash + m + \mu\gamma_0)$, where
$\Dslash$ is the massless Dirac operator and the last term appears when the chemical potential $\mu$ is
non-zero. Now, $\Dslash$ satisfies $\gamma_5$-hermiticity: $\gamma_5 \Dslash \gamma_5 = \Dslash^\dagger$,
so that
\be
\gamma_5 (\Dslash + m + \mu\gamma_0) \gamma_5 = \Dslash^\dagger + m - \mu\gamma_0 = (\Dslash + m - \mu^*\gamma_0)^\dagger
\label{eq:gamma5}
\ee
Taking the determinant on both sides gives $\det(\Dslash + m + \mu\gamma_0) = \det^*(\Dslash + m - \mu^*\gamma_0)$,
which constrains the determinant to be real only if $\mu$ is zero or pure imaginary. In such case, an even number
of degenerate flavors (same $m$, same $\mu$) yields a non-negative factor in the integration measure over the gauge
fields, and standard Monte Carlo techniques apply. The same is true for pairs of flavors with opposite, real $\mu$, 
i.e. isospin chemical potential.

In the general case, the determinant may be complex, and in fact it {\em must} be complex to produce the expected physics.
This can be seen by considering the free energy of a static color charge or anti-charge, respectively related to the 
expectation value of the Polyakov loop or its adjoint. Denoting by $d\varpi$ the integration measure which includes the
determinant, one sees that
\bea
\mbox{
$\langle {\rm Tr~Polyakov}~\: \rangle = \exp(-\frac{1}{T} F_{q}) = \int ~ {\rm Re(Polyakov)}\times {\rm Re}(d\varpi) {-} {\rm Im(Polyakov)}\times {\rm Im}(d\varpi)$} \\
\mbox{
$\langle {\rm Tr~Polyakov}^{{*}} \rangle = \exp(-\frac{1}{T} F_{{\bar{q}}}) 
= \int ~ {\rm Re(Polyakov)}\times {\rm Re}(d\varpi) {+} {\rm Im(Polyakov)}\times {\rm Im}(d\varpi)$}
\eea
Different free energies $F_q$ and $F_{\bar{q}}$, as happens when a chemical potential favors charge over anti-charge,
can only be obtained if ${\rm Im}(d\varpi) \neq 0$, i.e. with a complex measure\footnote{In the $SU(2), N_f=2$ case, the
square of the determinant remains real positive even when $\mu\neq 0$. But the baryonic chemical potential can be turned
into an isospin chemical potential by a redefinition of the quark fields.}.

A corollary of the above statement is that any Monte Carlo ensemble (which is sampled using a real non-negative measure)
has average baryon number zero (or pure imaginary).
 So the direct sampling of a finite-density ensemble is not possible. Three main approaches have been pursued towards 
 circumventing this difficulty: reweighting, Taylor expansion and analytic continuation from imaginary $\mu$.
I will review them in succession, emphasizing their application to the determination of the pseudo-critical temperature
$T_c(\mu)$ and of the QCD critical point.


\section{Reweighting}


\subsection{General results}
\label{subsec:sign}

\begin{figure}
\centerline{
\includegraphics[width=0.65\textwidth]{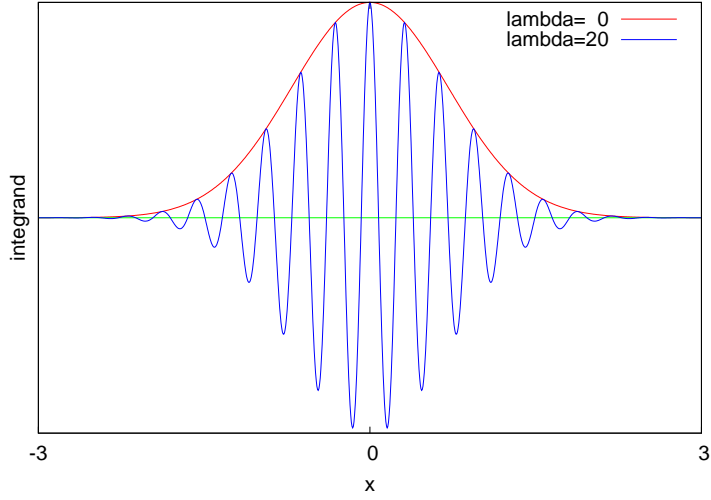}
}
\caption{Toy example of an oscillatory integrand: ``every $x$ is important''.}
\label{fig:2d_oscillatory}
\end{figure}

Let me first illustrate the problem in a toy model. Consider the ``partition function''
$Z(\lambda) \equiv \int_{-\infty}^{+\infty} dx \exp(-x^2 + i\lambda x)$. Since $Z(\lambda)$
is real, we can focus on the real part of the integrand, shown Fig.~\ref{fig:2d_oscillatory}.
While the important values of $x$ are clearly near zero when $\lambda=0$, this is no longer
true when $\lambda\neq 0$. Large cancellations take place, and integration far into the tail
of the distribution is needed to obtain the analytic result $Z(\lambda)/Z(0) \!=\! \exp(-\lambda^2/4)$.
The size of the ``important'' integration region is governed by $\lambda$, not by the width of the $\lambda=0$ Gaussian.
The situation is similar in QCD, where configurations suppressed by ${\cal O}(\exp(-{\rm Volume}))$ must
be properly summed over, as stressed most forcefully by Splittorff and collaborators~\cite{Osborn:2008eg}.
Loosely speaking, ``every configuration is important'', and it is not even clear how to sample them.

In general, one may ask the question:
given a real, oscillatory integrand $f(x)$, what is the optimal weight $g(x) \geq 0$ which should be used
for sampling? Uniform sampling is not the answer, since it is clearly wasteful to sample regions where $|f(x)|$ 
is zero or very small.
A precise answer can be obtained by considering how one forms the expectation value $\langle W \rangle_f$
of a general observable $W$ in the desired, target ensemble with partition function $Z_f \!=\! \int dx~f(x)$:
\be
\langle W \rangle_f = \frac{\int dx~W(x) f(x)}{\int dx~f(x)} = 
 \frac{\int dx~W(x) \frac{f(x)}{g(x)} g(x)}{\int dx~\frac{f(x)}{g(x)} g(x)} 
= \frac{\langle W \frac{f}{g} \rangle_g}{\langle \frac{f}{g} \rangle_g}
\label{eq:reweight}
\ee
This is the strategy of {\em reweighting}: successive measurements of $W$, obtained from
ordinary Monte Carlo sampling of the auxiliary partition function $Z_g \!=\! \int dx~g(x), ~ g(x) \geq 0$,
are given a varying, oscillatory weight $f/g$ in the ensemble average.
The denominator $\langle \frac{f}{g} \rangle_g = Z_f / Z_g$ is called the ``average sign''.
As we will see shortly, it becomes exponentially small as the volume is increased. In addition,
the relative error on the average sign propagates to every observable $W$. Therefore, $g(x)$ should
be chosen so as to minimize the relative variance of $f/g$. In the limit where the average sign
tends to zero, the solution is $g(x) = |f(x)|$ (up to an arbitrary multiplicative constant)~\cite{deForcrand:2002pa}.
Then, each measurement of the reweighting factor $f/g$ gives $\pm 1$, and
$Z_g$ is often called the ``sign quenched'' ensemble.

Since the average sign is a ratio of two partition functions, it can be rewritten as
\be
\left\langle \frac{f}{g} \right\rangle_g = \frac{Z_f}{Z_g} \underset{V\to\infty}{=} \exp\left(-\frac{V}{T} \Delta f(T,\lambda)\right)
\label{eq:Delta_f}
\ee
for a system of volume $V$ at temperature $T$, where $\Delta f$ is the free energy density difference
between the two ensembles, which depends on the temperature and the couplings of the theory ($\mu$ for QCD). 
This makes clear the dependence of the average sign on the volume. To maintain statistical accuracy, the
number of independent measurements must grow as $\exp(+2 \frac{V}{T} \Delta f(T,\lambda))$, i.e., exponentially fast
with $V$.

\subsection{Reweighting for QCD}
\begin{figure}
\centerline{
\includegraphics[width=0.50\textwidth]{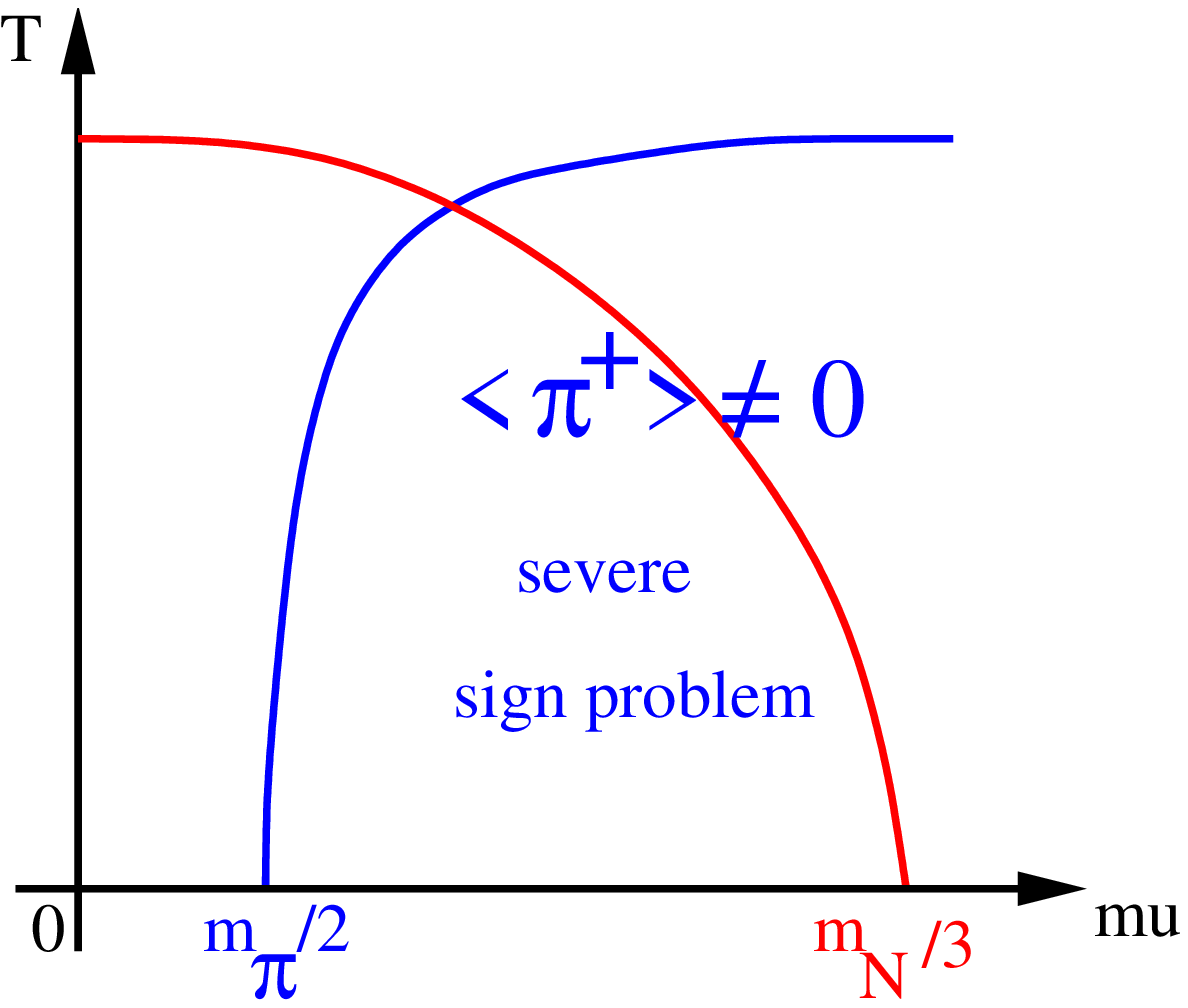}
\includegraphics[width=0.50\textwidth]{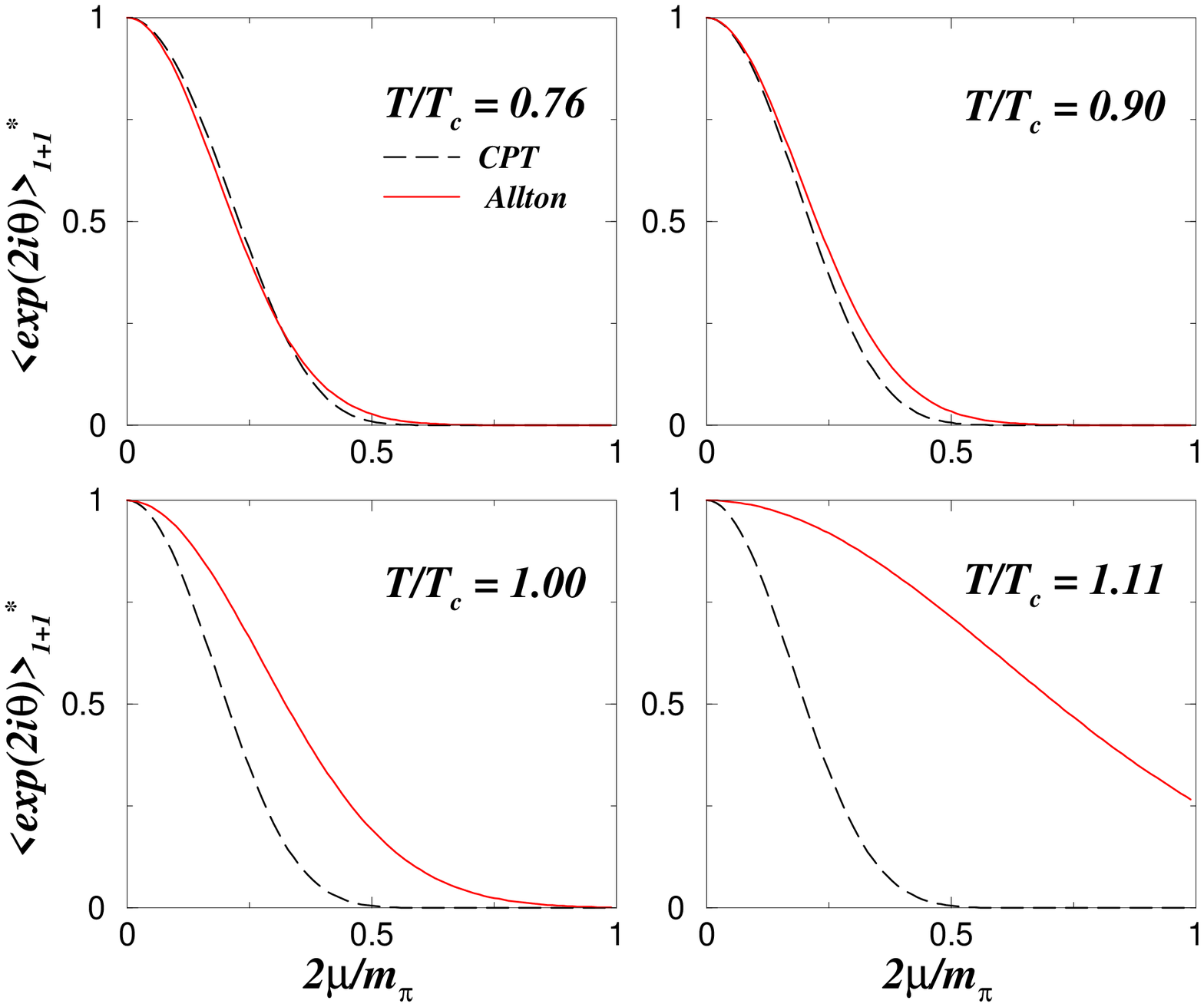}
}
\caption{{\em Left}: Sketch of the QCD pseudo-critical line $T_c(\mu)$ (in red), starting from $\sim m_N/3$ at $T=0$,
superimposed with the phase transition line (in blue) of the phase-quenched theory (alias isospin chemical potential),
starting from $m_\pi/2$ at $T=0$. 
Bose-Einstein condensation of charged pions in the phase-quenched ensemble causes a severe sign problem.
{\em Right}: Comparison of values of the ``average phase factor'' $\langle \exp(2 i \theta) \rangle$, measured in lattice simulations and predicted by one-loop chiral perturbation theory~\cite{Splittorff:2007zh}. Good agreement persists up to $T/T_c \sim 0.90$. 
}
\label{fig:severe}
\end{figure}

For QCD, the optimal choice of Monte Carlo probability is therefore $|{\rm Re}(\det(\mu)^{N_f})|$~\cite{deForcrand:2002pa,Hsu:2010zz}.
However, this expression cannot be recast, as customary, into a Gaussian integral for further stochastic estimation. 
This causes
a considerable ${\cal O}(V^2)$ computational overhead. A more appropriate choice is the ``phase quenched'' ensemble
with probability $|\det(\mu)^{N_f}|$. Given eq.~(\ref{eq:gamma5}), this can be rewritten as
$\det(+\mu)^{N_f/2} \det(-\mu)^{N_f/2}$, corresponding to an isospin chemical potential applied to $N_f/2$ pairs 
of flavors $(i,j)$. This latter form can be readily recast into a Gaussian integral and sampled with the usual Rational Hybrid Monte Carlo. However, the chemical potential is now coupled to all $q_i\bar{q_j}$ mesons. The lightest among those,
the charged pions $u\gamma_5 \bar{d}$ (using $N_f=2$ notation), undergo Bose-Einstein condensation when $\mu > \mu_c(T)$,
with $\mu_c(T=0) = m_\pi/2$, as illustrated Fig.~\ref{fig:severe} {\em left}. Then, the physics of the auxiliary Monte Carlo ensemble differs qualitatively from that of
the target baryonic-$\mu$ ensemble, which causes $\Delta f$ in eq.~(\ref{eq:Delta_f}) to become large: the sign problem
becomes {\em severe}.

Remarkably, the average sign is mostly determined by the physics of pions, for which chiral perturbation theory,
or even random matrix theory,
provides an accurate analytic description provided $T \ll m_\pi$ and $\mu \ll m_\rho/2$.
A convenient observable to study is the ``average phase factor'' $\langle \exp(2 i \theta) \rangle = \frac{Z(+\mu,+\mu)}{Z(+\mu,-\mu)} = \langle \frac{\det(\mu)^2}{|\det(\mu)|^2} \rangle_{||}$,
where $Z(+\mu,+\mu)$ is the target ensemble with chemical potential $\mu$, $Z(+\mu,-\mu)$ is the auxiliary
Monte Carlo ensemble with isospin chemical potential, and $\langle .. \rangle_{||}$ is an expectation value
with respect to the latter. The observable $\langle \exp(2 i \theta) \rangle$ measures the effect of changing the chemical
potential from $+\mu$ to $-\mu$ for half of the quark flavors, or equivalently the corresponding fermion boundary
conditions in Euclidean time. It is thus ultra-violet finite and can be estimated using continuum chiral
perturbation theory.
Comparison between analytic and numerical lattice QCD results shows good agreement, even at rather high
temperatures not far from $T_c$~\cite{Splittorff:2007zh}: see Fig.~\ref{fig:severe} {\em right}.

Pion condensation is a consequence of choosing to sample a Monte Carlo ensemble with isospin chemical potential.
One may wonder if the severe sign problem could not be avoided with another choice of Monte Carlo ensemble, e.g.,
$\mu=0$. However, 
it is still because of the {\em phase} of the reweighting factor that pion condensation for $\mu > m_\pi/2$ does not occur.
Additional fluctuations in its magnitude only cause 
the evaluation of the average sign to be even more noisy, and the sign problem even more severe.
In addition, a pernicious effect occurs: the distribution of the reweighting factor becomes broader and non-Gaussian, which
makes the analysis of the statistical error less reliable. Moreover, the necessarily finite Monte Carlo sample may
contain zero configurations in the region which is most important for the target ensemble. This leads to wrong results,
because the underestimated statistical error does not reflect the large deviation from the correct answer.
Failure of reweighting may go undetected, as in the unfortunate ``Glasgow method''~\cite{Barbour:1986jf}.

\begin{figure}[t]
\vspace*{-1.2cm}
\centerline{
\includegraphics[width=5.0cm]{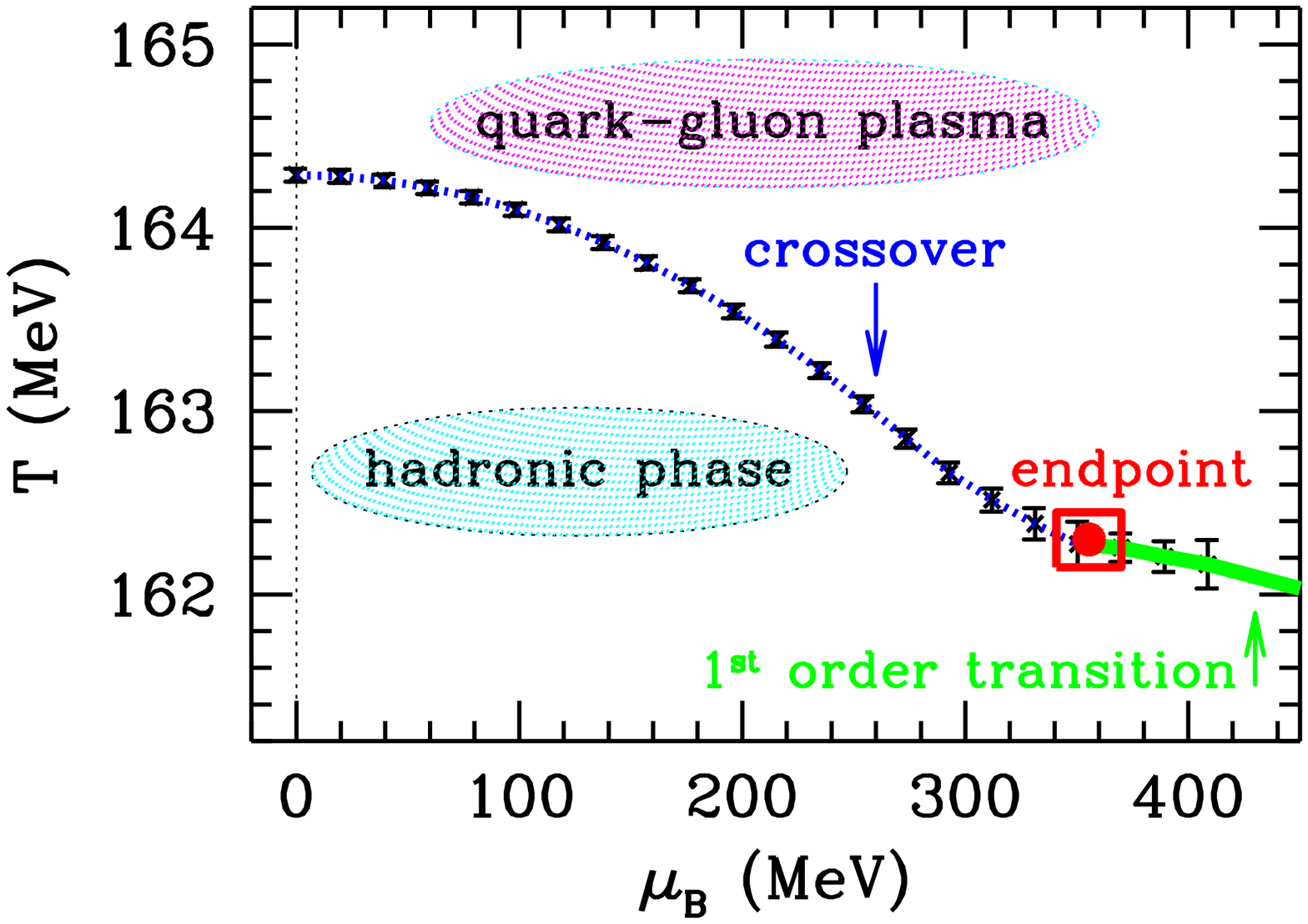}
\hspace*{0.1cm}
\includegraphics[width=5.0cm]{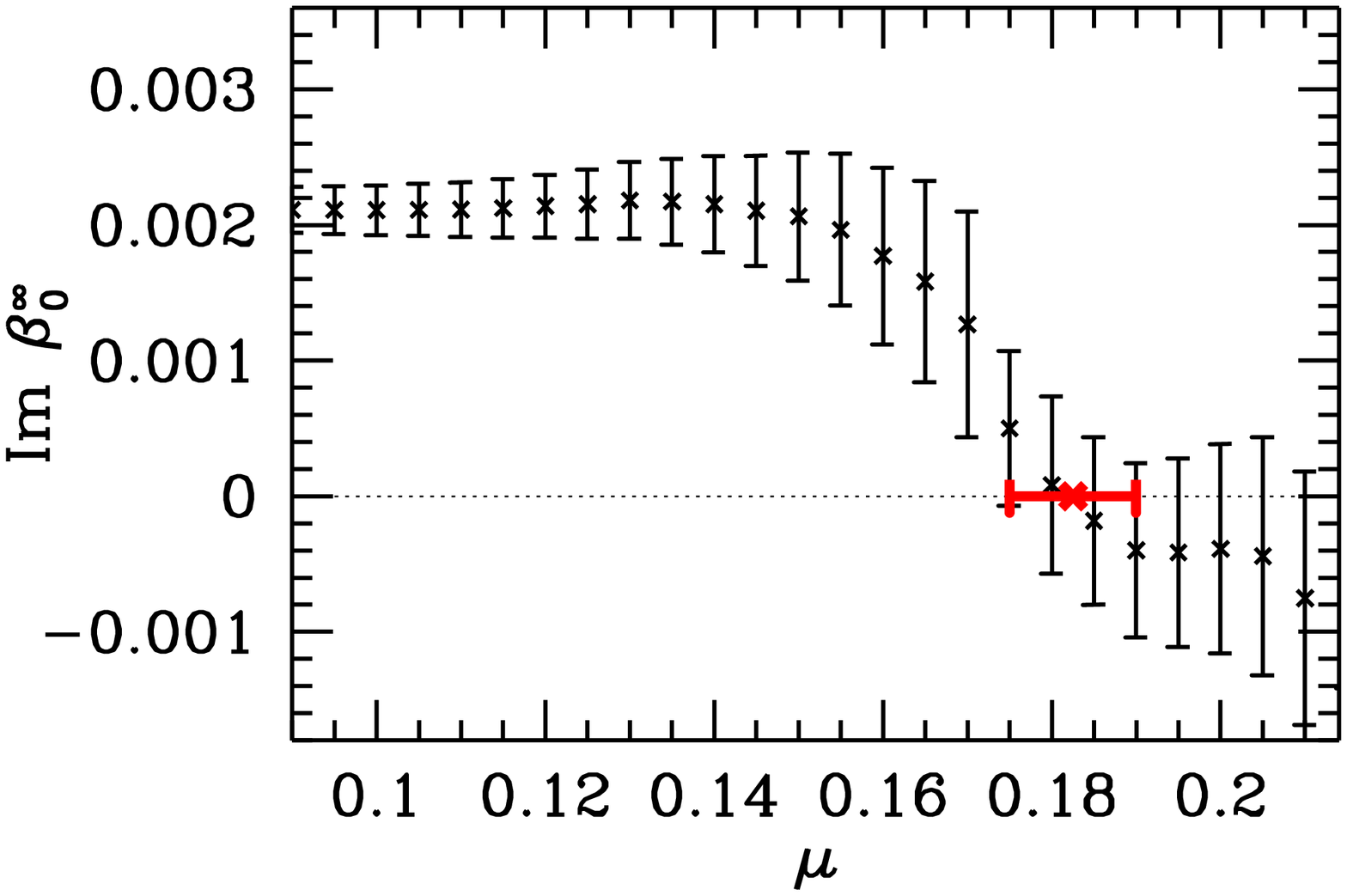}
\put(-261.5,73){{\color{blue}$\bullet$}}
\put(-258,35){$\longrightarrow$}
\put(-238,35){$\longrightarrow$}
\put(-218,35){$\longrightarrow$}
\put(-238,30){\tiny Glasgow}
\hspace*{0.1cm}
\includegraphics[width=5.0cm,height=4.75cm]{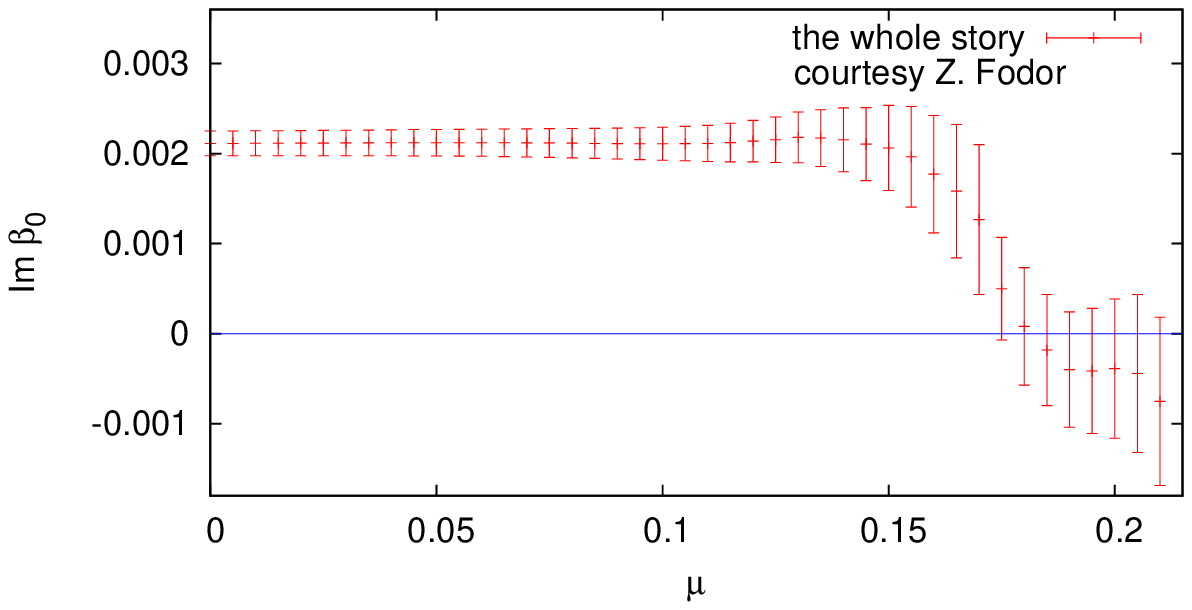}
}
\caption{{\em Left}: QCD phase diagram from \cite{Fodor:2004nz} obtained by combined reweighting in $\mu$ and $\beta$ of
the $\mu=0, \beta=\beta_c$ ensemble (blue dot). {\em Middle}: corresponding smallest Lee-Yang zero imaginary part (related to 
the inverse of the specific heat) extrapolated to the thermodynamic limit. {\em Right}: full data illustrating
the insensitivity to $\mu$ followed by an abrupt change, courtesy of Z.~Fodor.}
\label{fig:FK}
\end{figure}

In spite of these difficulties, reweighting has produced a landmark result~\cite{Fodor:2004nz}, with a determination of the pseudo-critical
temperature $T_c(\mu)$ and of a critical point (see Fig.~\ref{fig:FK}) for QCD with physical quark masses, on an $N_t=4$
(4 time-slices, $a\sim 0.3$ fm) lattice. The Monte Carlo ensemble chosen, $(\mu=0,\beta=\beta_c(\mu=0))$, was sub-optimal,
yet vastly superior to the earlier ``Glasgow method'' which kept $\beta$ fixed~\cite{Barbour:1986jf}.
Still, one may question whether the statistical error, which one would expect to grow exponentially with $\mu^2$,
is reliably evaluated. Doubts are fueled by the observation of \cite{Splittorff:2005wc} that the critical point is located near
the estimated boundary to the pion-condensed phase (Fig.~\ref{fig:severe} {\em left}), and that the system appears insensitive to
$\mu$ until very close to the critical point (Fig.~\ref{fig:FK}, right). Six years onward, this issue is still not settled,
in spite of the authors' own efforts to devise a more reliable error estimation~\cite{Csikor:2004ik}, 
and doubts about the determination of the critical point linger on.

\begin{figure}
\centerline{
\includegraphics[width=5.2cm]{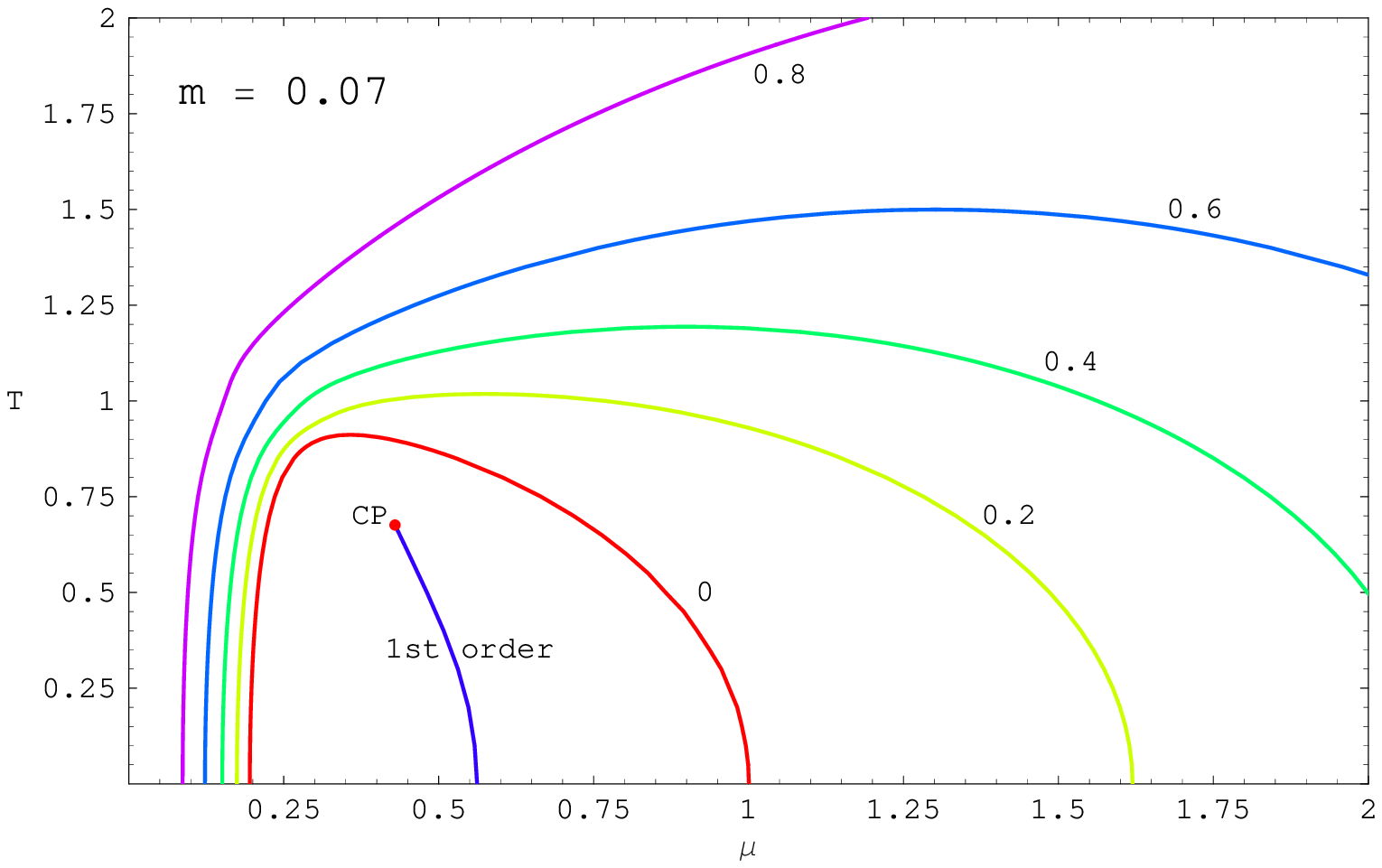}
\hspace*{2.5cm}
\includegraphics[width=4.5cm]{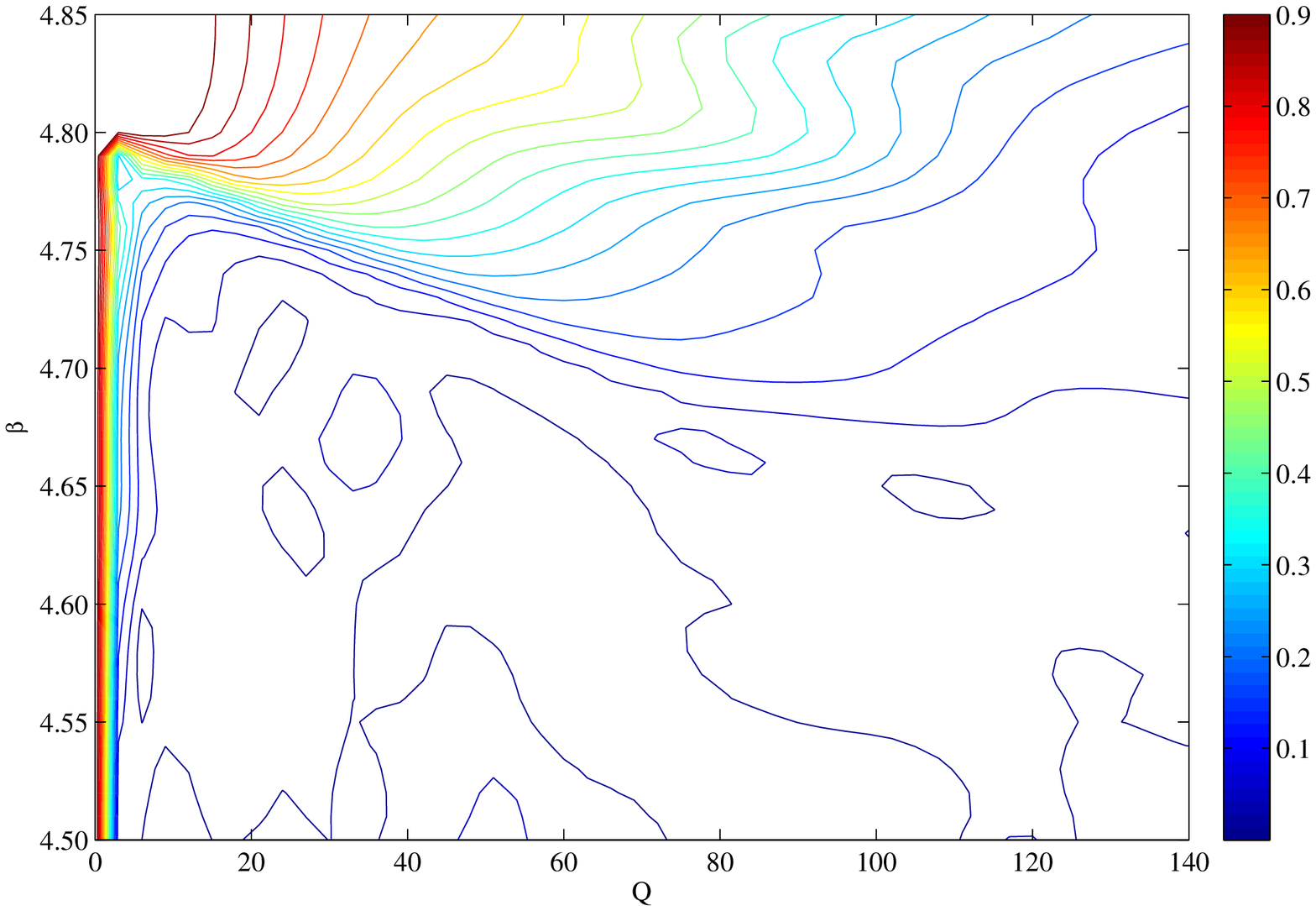}
}
\caption{Isolines of the average sign in the ($\mu,T$) plane for a random matrix model~\cite{Han:2008xj} 
and in the (density,~$\beta$) plane for $N_f=8$ simulations~\cite{deForcrand:2007uz}.}
\label{fig:isolines}
\end{figure}

Progress is continuing in the analytic estimation of the average sign. Recent work includes the influence of the topological
sector at finite $\mu$~\cite{Bloch:2008cf}, and the determination of the {\em distribution} of the phase of the determinant, as well as its correlation with various observables like the baryon number~\cite{Lombardo:2009aw}. Even a random matrix model with a critical 
point~\cite{Han:2008xj} gives a description of the sign problem roughly consistent with numerical reweighting~\cite{deForcrand:2007uz}:
compare Fig.~\ref{fig:isolines} {\em left} and {\em right}. 
One should note that pion interactions do not play a major role in the determination of the average sign. On the other hand,
taking the baryons into account improves the description (and turns out to make the sign problem less severe). This can be accomplished by describing
the two systems (baryonic-$\mu$ and isospin-$\mu$) by a non-interacting hadron resonance gas. A fit of lattice simulation
results to such an ansatz~\cite{D'Elia:2009tm} works well. Interestingly, one can then in turn {\em predict} the maximum baryon number
which can be included in a lattice by reweighting, for an average sign of, say, $0.1$ or greater. For practical lattice
sizes, this number is ${\cal O}(10)$ (and decreases for lighter quarks), which is barely sufficient for a statistical
treatment.
   

\section{Taylor expansion}
\label{sec:Taylor}

\begin{figure}
\centerline{
\includegraphics[width=4.80cm]{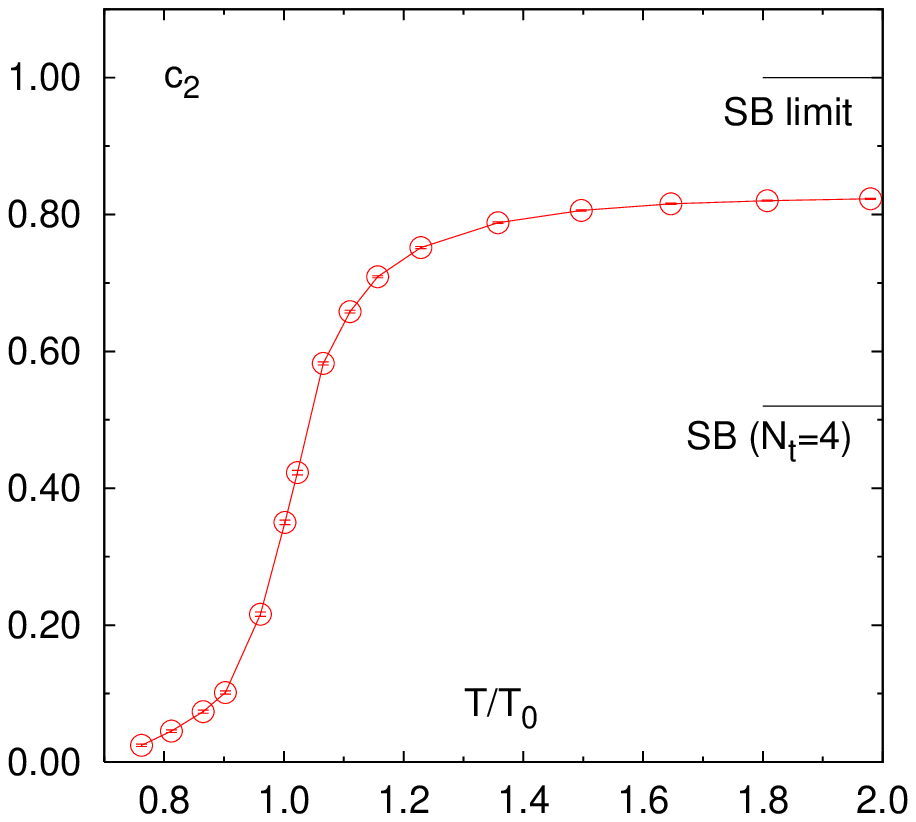}
\includegraphics[width=4.80cm]{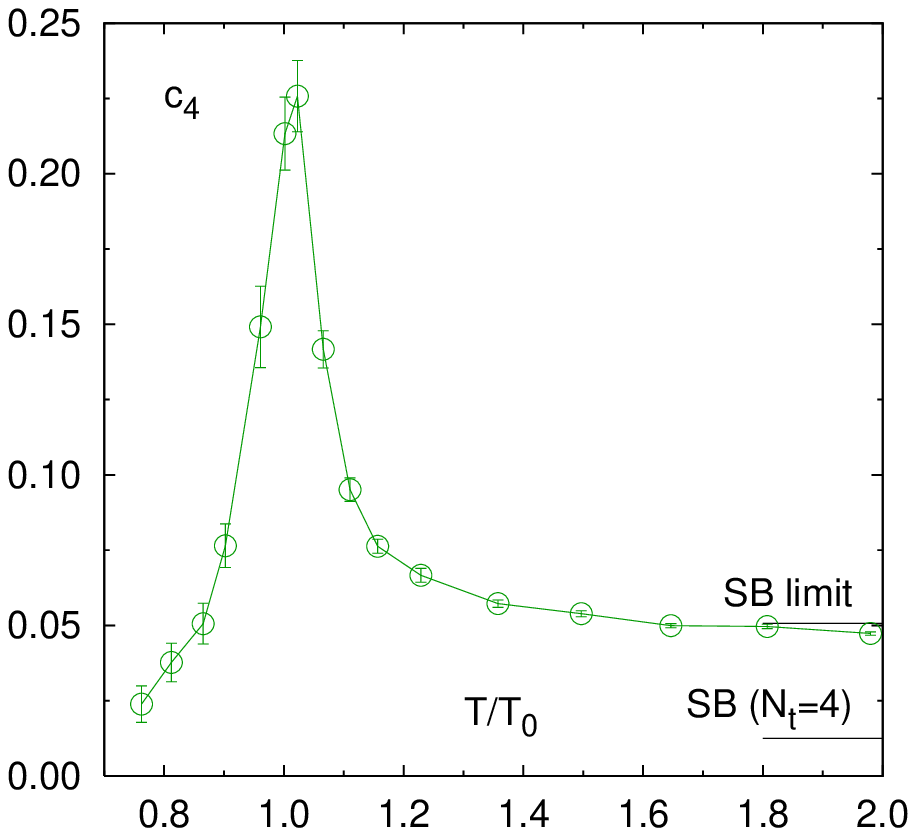}
\includegraphics[width=4.80cm]{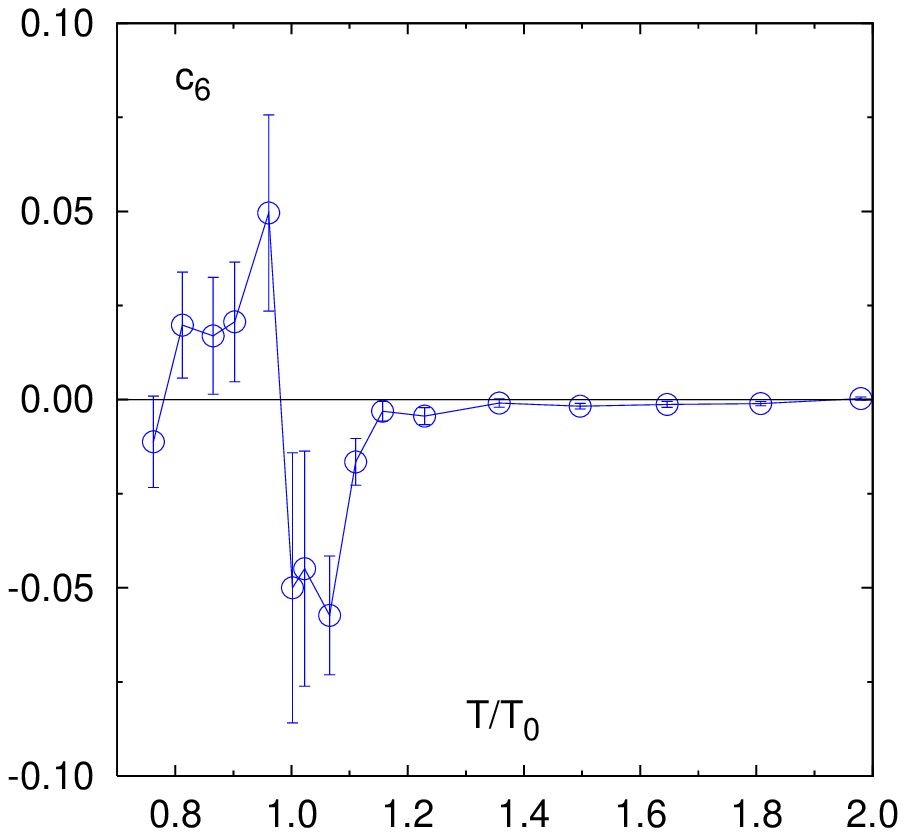}
}
\caption{First three coefficients in the Taylor expansion of the QCD pressure eq.~(\protect\ref{eq:pressure}) versus $T/T_c$~\cite{Allton:2005gk}.}
\label{fig:Taylor}
\end{figure}

As we have seen, reweighting is limited to small volumes, and its breakdown is difficult to detect.
It may be more useful and efficient to try and determine, in the thermodynamic limit, the first few Taylor coefficients
in the expansion of an arbitrary observable in powers of $\mu/T$ about $\mu=0$. 
In particular, one may consider the pressure $P(T,\mu)$, since all thermodynamic properties can be extracted from its
derivatives. Defining $\Delta P(T,\mu) \equiv P(T,\mu) - P(T,\mu=0)$, one expands
\bea
\frac{\Delta P(T,\mu)}{T^4} = \sum_{k=1}^\infty {c_{2k}(T)} \left( \frac{\mu}{T} \right)^{2k} 
\label{eq:pressure} \\
{c_{2k}} = \langle {\rm Tr}({\rm degree}~ 2k ~{\rm polynomial ~ in ~} \Dslash^{-1}, \frac{\partial\Dslash}{\partial\mu})  \rangle_{\bf\mu=0}
\label{eq:c_2k}
\eea
where the Taylor coefficients $c_{2k}$ can be expressed as expectation values of traces of matrix polynomials 
in the $\mu=0$ ensemble. 
The trace of each monomial can then be estimated by the standard stochastic averaging over ``noise vectors''.
This strategy looks straightforward, and indeed works well at low orders: see Fig.~\ref{fig:Taylor}~\cite{Allton:2005gk}. 
Notice however how the
statistical errors grow with the Taylor order. Since one must go to higher order as $\mu$ is increased, to keep 
the truncation error of the Taylor expansion under control, an essential practical question is: how does the work increase
with the order $k$ ?

The answer has several parts, and a full complexity analysis has not been carried out yet: \\
$\bullet$ The number of terms in the degree $2k$ polynomial grows approximately as $6^{2k}$~\cite{Schmidt_private}. \\
$\bullet$ The Taylor coefficient $c_{2k}$ has a finite thermodynamic limit, but the monomials in eq.~(\ref{eq:c_2k}) grow as $V^{2k}$,
implying large cancellations, which can only be controlled by a large (exponential in $k$) increase with $V$ in the number of noise vectors.
This is where the complexity $\sim \exp(V)$ of the reweighting approach resurfaces. After all, the underlying strategy
is an extrapolation in $\mu$ at fixed $\beta$, much like the Glasgow method. \\
$\bullet$ The distribution of the values whose average yields $c_{2k}$ is less and less Gaussian as $k$ increases. \\
$\bullet$ Finite-size effects grow with $k$, since $c_{2k}$ is analogous to a $2k$-point function. \\
These considerations should lead us to expect steady, but slow progress in the expansion to higher order.
In my opinion, increasing $k$ by one requires increasing computer resources by well over two orders of magnitude.
The current state of the art is $k_{\rm max}=4$ ($8^{th}$ order expansion) on an $N_t=6$ lattice~\cite{Gavai:2008zr}.

In this situation, it is worth exploring alternative methods to obtain the Taylor coefficients. These are based on
simulations at imaginary chemical potential.

\section{Imaginary $\mu$}

\begin{figure}
\centerline{
\includegraphics[width=5.50cm]{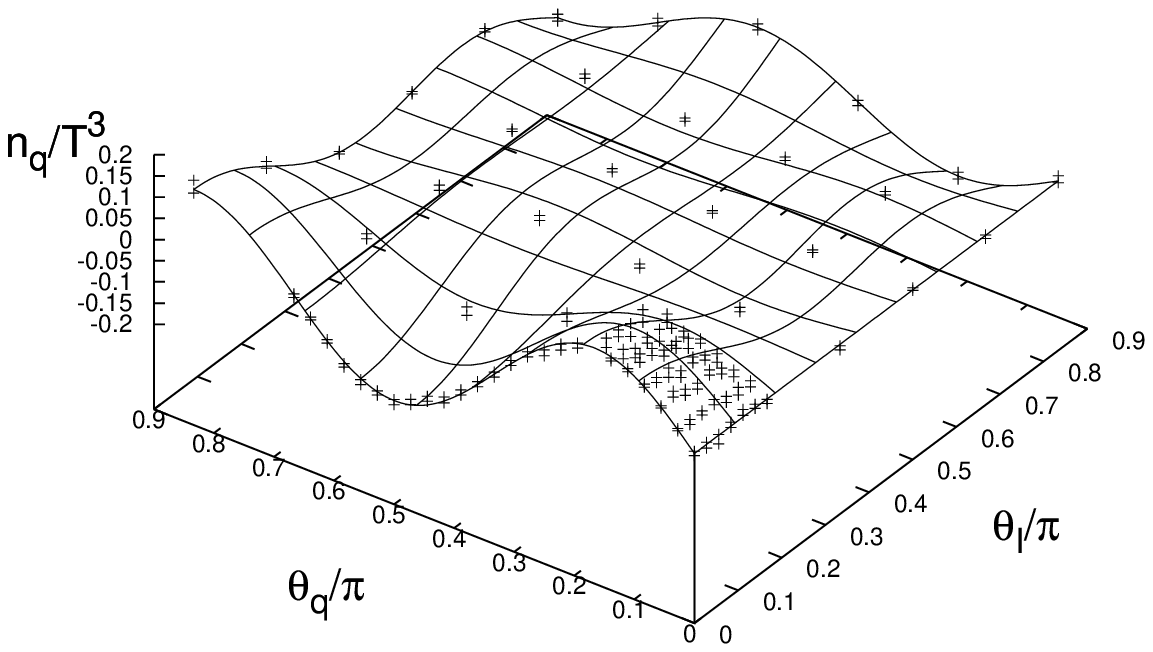}
\hspace*{1.5cm}
\includegraphics[width=5.50cm]{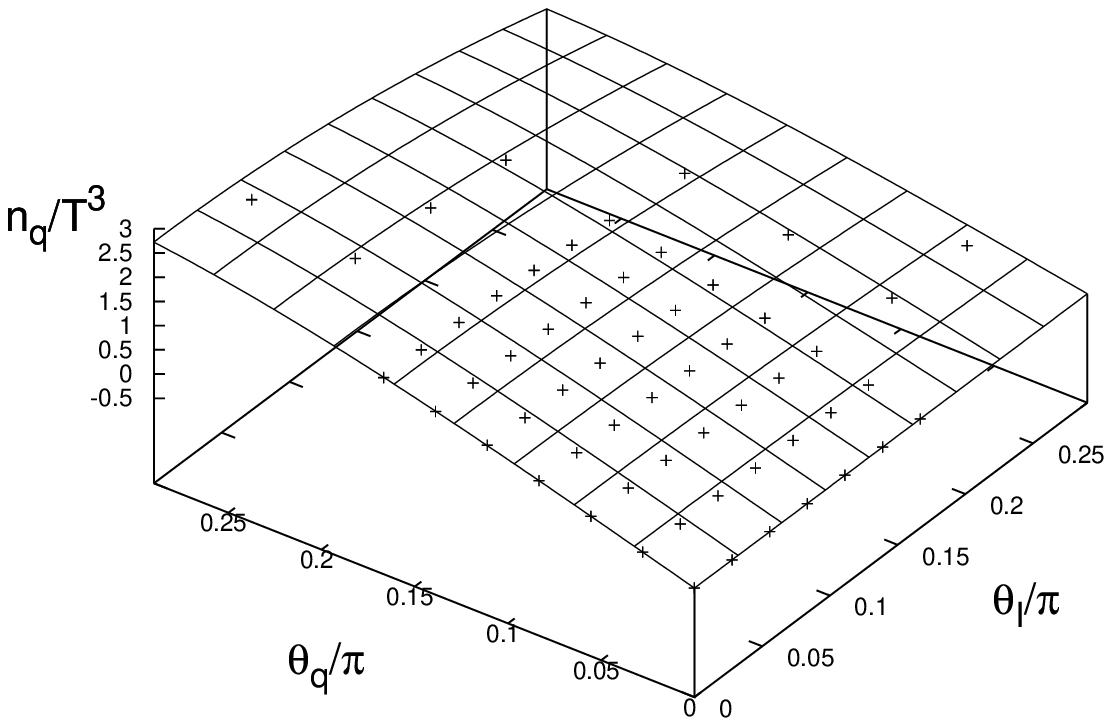}
}
\caption{Imaginary quark density as a function of imaginary quark and isospin chemical potentials, for temperatures $0.9 T_c$ ({\em left})
and $1.25 T_c$ ({\em right}), from \cite{D'Elia:2009tm}. The surfaces represent hadron resonance gas and polynomial fits, respectively.}
\label{fig:DElia_Taylor}
\end{figure}

The strategy is simple: perform independent simulations at different values of the imaginary chemical potential
$\mu = i \mu_i$, fit the results with an ansatz, and analytically continue the ansatz to real $\mu$.
If the ansatz is polynomial, the fit parameters are the usual Taylor coefficients.
Although this approach has been used mostly to determine the pseudo-critical temperature $T_c(\mu)$, it has
also been applied to the pressure, yielding the same Taylor coefficients
$c_{2k}$ as in eq.~(\ref{eq:c_2k}). A recent study~\cite{D'Elia:2009tm} is illustrated Fig.~\ref{fig:DElia_Taylor}.
At low temperature, the pressure is best described by a hadron resonance gas ansatz. For $T \geq 0.95 T_c$,
this ansatz becomes poor, and a better description is obtained by a Taylor expansion, which is sensitive to $c_6$.
Similar observations have been made in Ref.~\cite{Takaishi:2010kc} on a smaller lattice. A technical difference is that Ref.~\cite{D'Elia:2009tm} measures only the quark density, i.e. the first derivative of the pressure, as a function
of imaginary quark and isospin chemical potentials both, while Ref.~\cite{Takaishi:2010kc} measures all derivatives
in $\mu_u, \mu_d$ up to 4th order, but as a function of quark chemical potential only.
It would make sense to marry the two approaches: derivatives up to 4th order are easy to compute, and imaginary 
isospin chemical potential straightforward to implement.
Another important technical issue should be addressed: how to choose the simulated values of imaginary chemical 
potential and the statistics for each value, so as to maximize the accuracy on a given set of Taylor coefficients?
Larger values of $\mu_i$ increase the sensitivity to the desired higher-order terms, but also the truncation error in the
fitted Taylor polynomial.

\section{Results: $T_c(\mu)$}
\label{sec:Tc}

\begin{figure}[t]
\vspace*{-2.4cm}
\centerline{
\includegraphics[width=5.5cm]{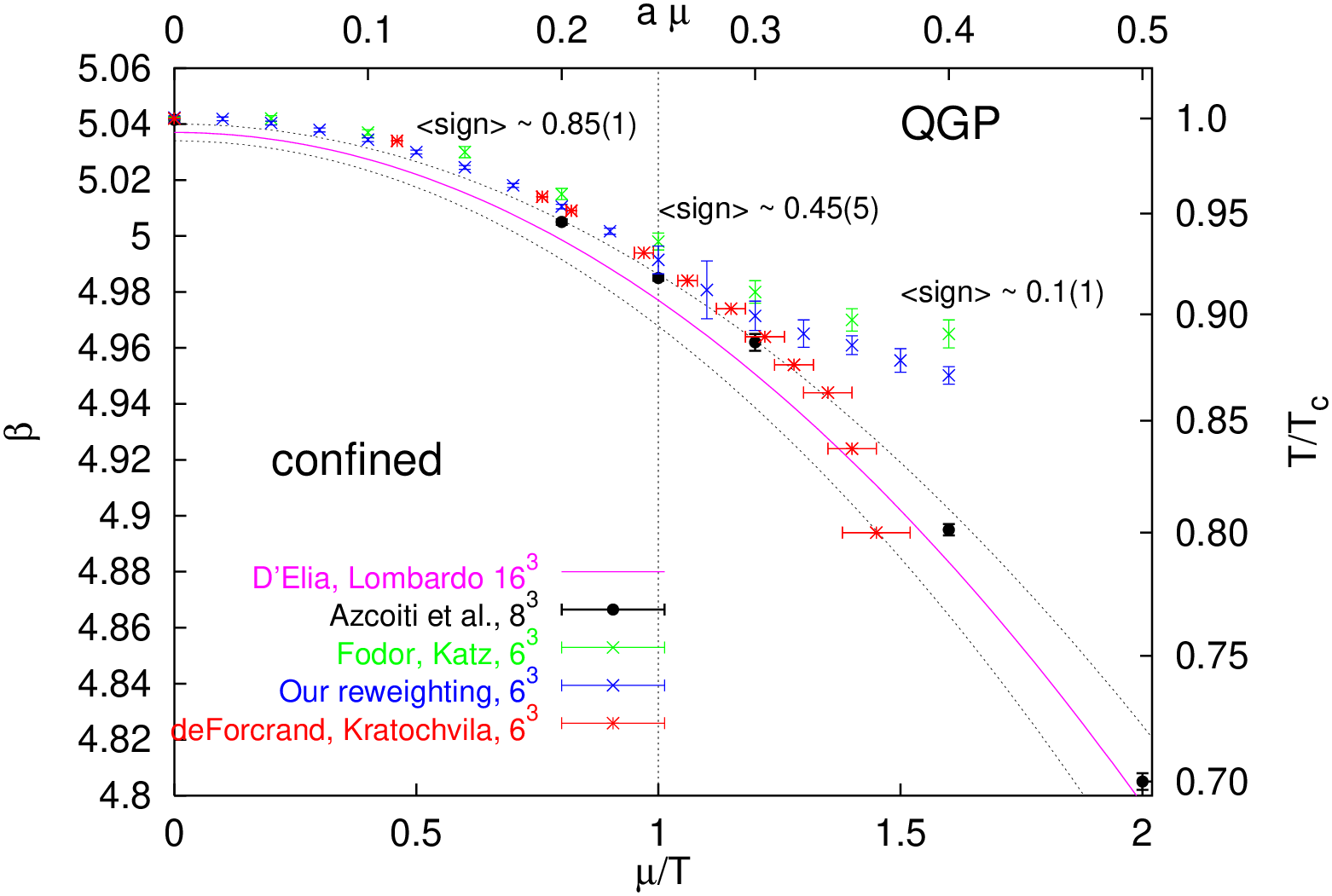}
\put(-195,52){\tiny {imaginary $\mu$}}
\put(-205,46){\tiny {2 param. imag. $\mu$}}
\put(-205,40){\tiny {dble reweighting, LY zeros}}
\put(-205,34){\tiny {Same, susceptibilities}}
\put(-195,28){\tiny {canonical}}
\hspace*{2.3cm}
\includegraphics[width=6.5cm,height=4.9cm]{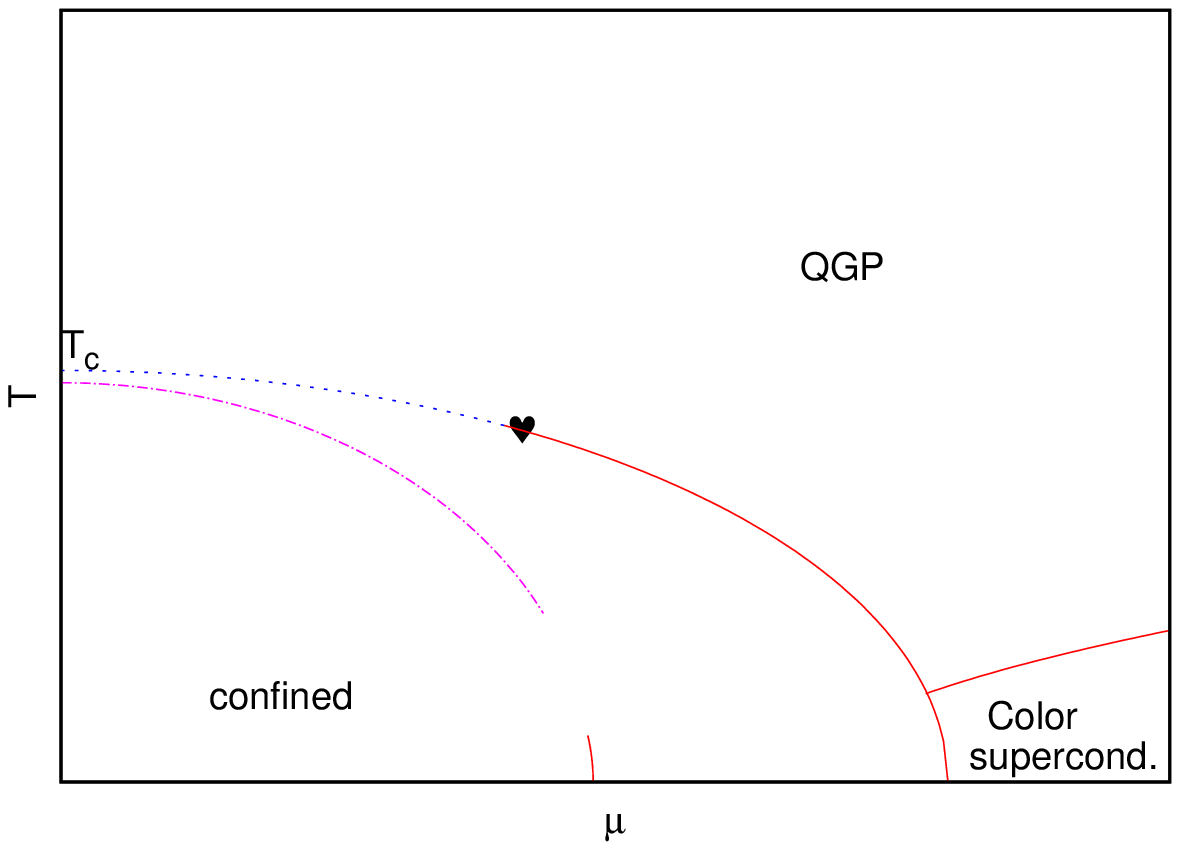}
\put(-130,60){\color{blue}\tiny freeze-out}
\put(-112,78){\color{red}\tiny crit. pt.}
}
\caption{({\em Left}) Pseudo-critical temperature determined by various approaches for the same lattice theory (4-flavor staggered quarks with mass $am=0.05$ on an $N_t=4$ lattice)~\cite{Kratochvila:2005mk}. All approaches agree for $\mu/T \lsim 1$.
({\em Right}) Effect of a small curvature of the pseudo-critical temperature $T_c(\mu)$: the distance from the putative
critical point to the freeze-out curve increases, and any signature of the critical point tends to be washed out.}
\label{fig:Slavo}
\end{figure}

The pseudo-critical temperature can be determined in a variety of ways. As illustrated Fig.~\ref{fig:Slavo} {\em left}, all 
determinations agree for small $\mu/T$. In particular, the curvature $t_2$ in the Taylor expansion
\be
\frac{T_c(\mu)}{T_c(\mu=0)} = 1 - \sum_{k=1} {\bf t_{2k}} \left( \frac{\mu}{\pi T} \right)^{2k}
\label{eq:T_c}
\ee
can be determined by the most economical method, which seems to be that of analytic continuation from imaginary $\mu$.
While in the past, coarse lattices with $N_t=4$ and 6 time-slices only have been considered, this year has seen remarkable
progress, with a first extrapolation to the continuum limit for physical quark masses~\cite{Endrodi}.
Using $N_t=4, 6, 8$ and 10 lattices and an imaginary-$\mu$ method, this study confirms earlier indications that the curvature of the pseudo-critical line decreases as $a\to 0$~\cite{deForcrand:2007rq} and is small~\cite{Philipsen:2008gf} compared to that of the freeze-out curve,
defined as the fireball temperature below which inelastic collisions stop taking place and the ``chemical'', hadronic composition
of the fireball decay products remains frozen.
Since one has a crossover at $\mu=0$, $T_c$ depends on the observable considered, and so does the curvature.
Still, all observables yield a curvature smaller than that of the freeze-out curve $t_2 \sim 2$~\cite{Cleymans:2006qe}.

This result is of phenomenological importance. As seen Fig.~\ref{fig:Slavo} {\em right}, a flatter pseudo-critical line $T_c(\mu)$ increases
the distance from the putative QCD critical point to the freeze-out curve, giving more time for a possible signature
of criticality to be washed out as the fireball expands before hadronization. Note, however, that the determination of
the freeze-out curve is still being debated (compare \cite{Cleymans:2006qe} with, e.g., \cite{Andronic:2009gj,Rafelski:2002ga}).
 
\begin{figure}
\centerline{
\includegraphics[scale=0.28,clip=true,trim=0mm 1mm 0mm 0mm]{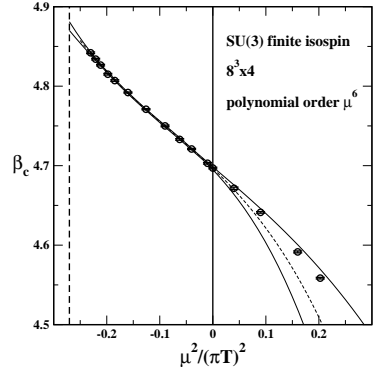}
}
\caption{Analytic continuation of the pseudo-critical line $T_c(\mu)$: for imaginary $\mu$ the Taylor series
is alternating, making the determination of the subleading Taylor coefficients difficult~\cite{Cea:2009ba}.}
\label{fig:Cea}
\end{figure}

The difficulties of determining subleading coefficients $t_{2k},~k>1$ in the expansion eq.~(\ref{eq:T_c}) has been considered 
in \cite{Cea:2009ba}, using the imaginary-$\mu$ approach in cases free of sign problem ($SU(2)$ and $SU(3)$ with isospin $\mu$) 
where the analytic continuation to real $\mu$ can be checked against a direct determination. For real $\mu$, the pseudo-critical line $T_c(\mu)$ bends down
more and more with increasing $\mu$. This indicates that the coefficients $t_{2k}$ are positive. Unfortunately,
for imaginary $\mu$ the Taylor series is then alternating. Successive contributions largely cancel each other,
and the $t_{2k}$'s are poorly determined, as illustrated Fig.~\ref{fig:Cea}. 
Technical issues like the best strategy for choosing simulation points and the best choice of fitting ansatz
are starting to be explored.

\section{Results: critical endpoint}

\begin{figure}
\centerline{
\includegraphics[width=0.43\textwidth]{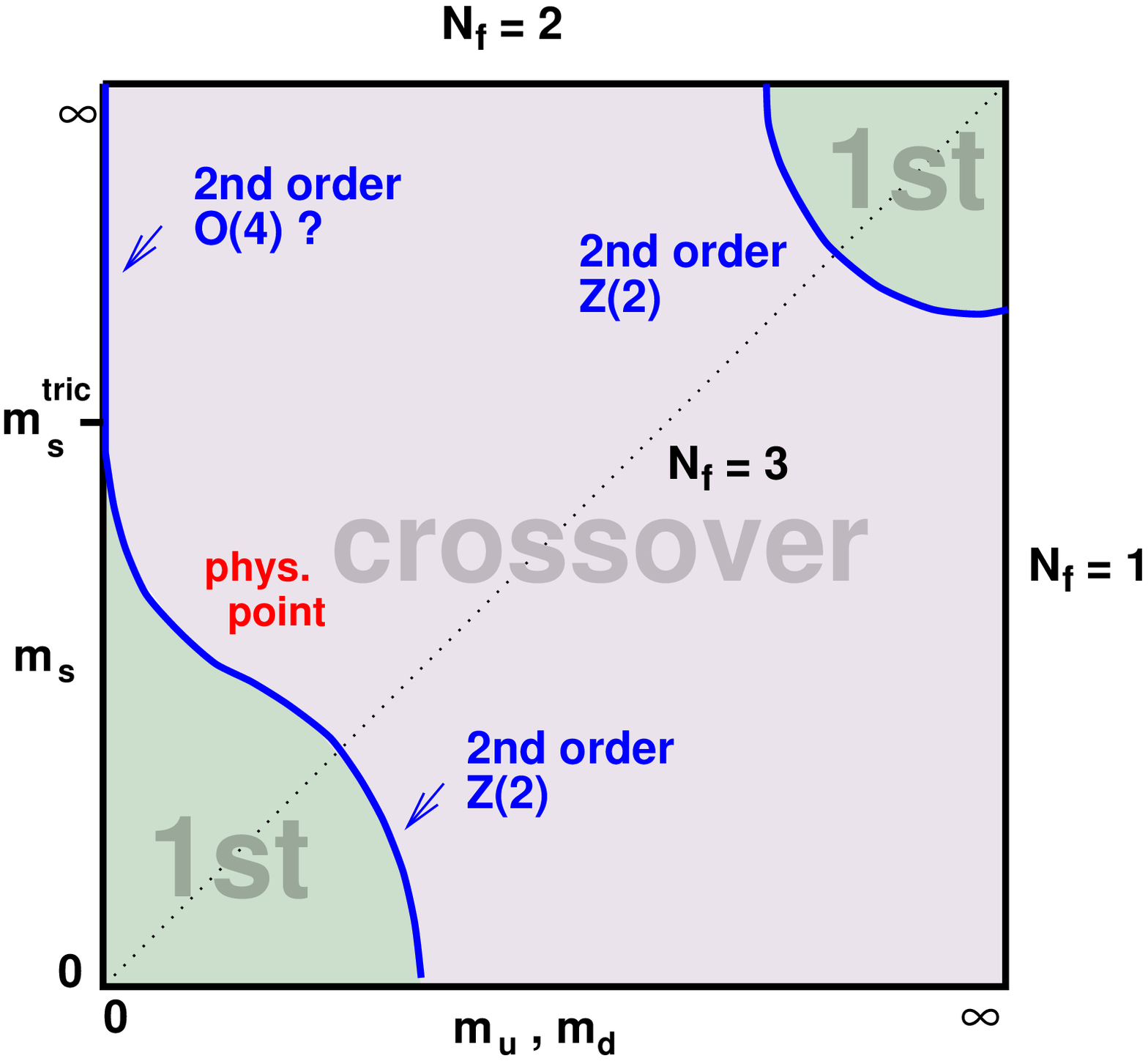}
\hspace*{-0.90cm}
\begin{minipage}{2.3cm}
\vspace*{-2.0cm}
{\large $\underbrace{\longrightarrow}_{{\rm if}~chiral~{\rm CEP}}$ }
\end{minipage}
\hspace*{-1.39cm}
\includegraphics[width=0.575\textwidth,clip=true,trim=4mm -15mm 0mm 0mm]{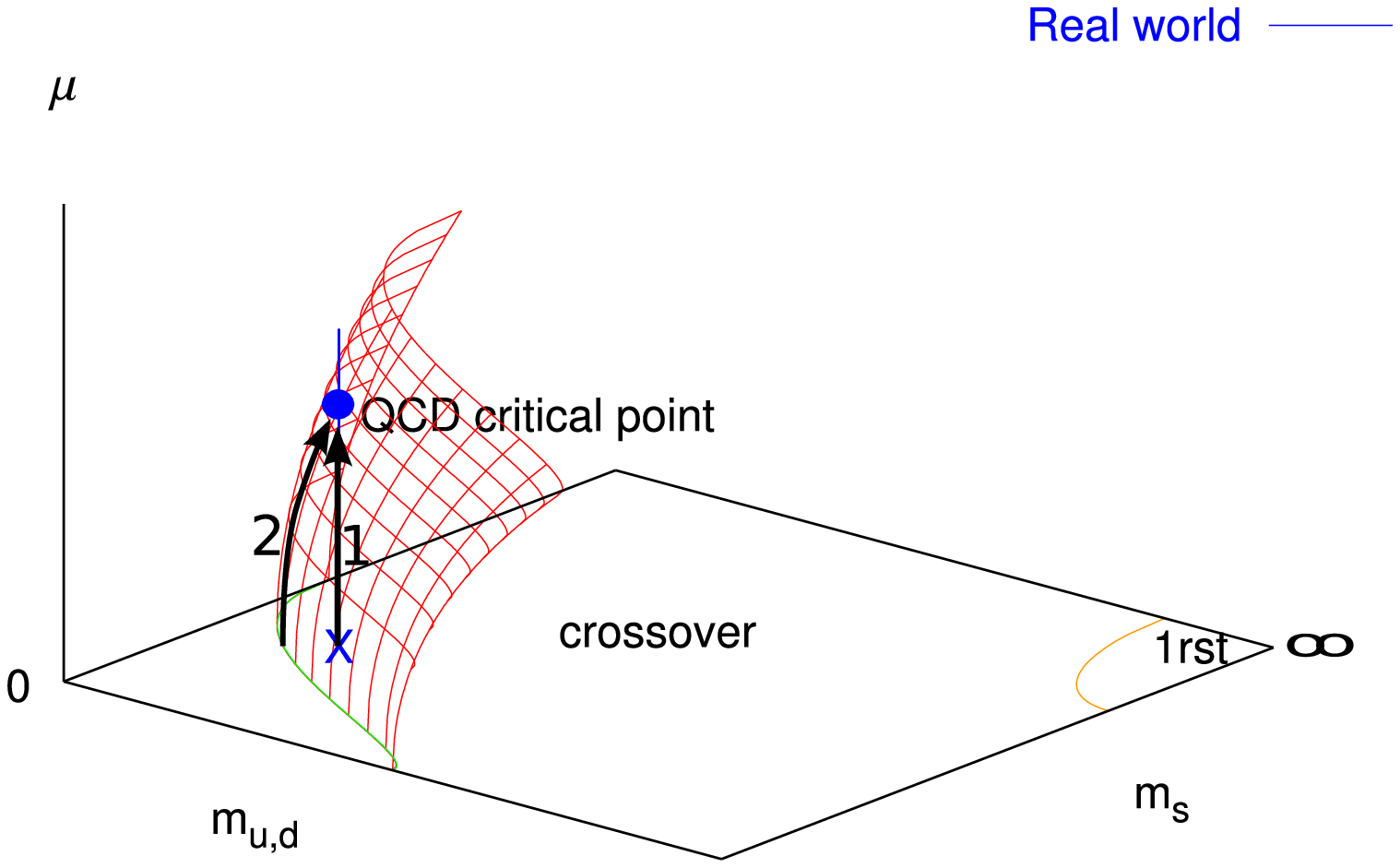}
}
\caption{({\em Left}) Order of the $\mu=0$ finite temperature transition as a function of the light and strange quark masses.
({\em Right}) Two strategies to reach the chiral critical point: (1) at fixed quark masses or (2) along the critical surface.}
\label{fig:critpt}
\end{figure}

Assume that the phase diagram of QCD features a critical point $(\mu_E,T_E)$, as marked by the star in Fig.~\ref{fig:phasediag}.
This critical point may or may not be {\em chiral}. A chiral critical point belongs to the chiral critical surface,
swept by the $\mu \!=\! 0$ chiral critical line in the lower left corner of Fig.~\ref{fig:critpt} {\em left} as $\mu$ is
turned on. A chiral critical point can be brought to $\mu \!\!=\!\! 0$ by tuning the quark masses, otherwise not.

A general strategy to locate the QCD critical point, chiral or not, is shown by the first arrow Fig.~\ref{fig:critpt} 
{\em right}: one looks for a singularity in $\mu$, keeping the quark masses fixed. All such approaches, except for the
reweighting study of Fodor and Katz~\cite{Fodor:2001au,Fodor:2004nz}, also keep the temperature fixed. One then has to address the delicate
question of how to correctly determine the temperature $T_E$.
For a chiral critical point, an alternative is to stay on the critical surface (second arrow Fig.~\ref{fig:critpt} {\em right}),
where the critical temperature is implicitly defined as a function of $\mu$ and the quark masses.
I review these two strategies in succession.

\subsection{Fixed mass, fixed temperature: ``effective radius of convergence''}

The determination of the Taylor expansion of the pressure, eq.~(\ref{eq:pressure}), provides in principle a simple way to estimate 
the location of the QCD critical point: the Taylor expansion will stop converging. The pole at $(\mu_E,T_E)$ in the second derivative $d^2P/d\mu^2\equiv \chi_q$ of the pressure will govern the divergence of the coefficients. In fact, there is another pole at $(-\mu_E,T_E)$, so that $d^2P/d\mu^2(\mu,T=T_E) \propto 1/(\mu_E-\mu) + 1/(\mu_E+\mu) \propto (\mu_E^2 - \mu^2)^{-1}$. It follows that
\be
\frac{\mu_E}{T_E} = \displaystyle\lim_{n\to\infty} \sqrt{\left|\frac{c_{2n}(T)}{c_{2n+2}(T)}\right|} ~~~ {\rm at} ~~~ T=T_E
\label{eq:radius}
\ee
Thus, some indication about the QCD critical point can perhaps be obtained for free from the first few Taylor coefficients.
Indeed, this approach has been followed by Karsch and collaborators~\cite{Allton:2005gk} and by Gavai and Gupta~\cite{Gavai:2008zr}.
The latter group has made strong statements, like 
``We find the radius of convergence of the series at various temperatures, and bound the location of the QCD critical point to be
$T_E/T_c \approx 0.94$ and $\mu_E/T_E < 0.6$'' (abstract Ref.~\cite{Gavai:2008zr}). Such strong statements force me to balance them with strong words of caution.

The first and obvious point is that eq.~(\ref{eq:radius}) concerns the $n\to\infty$ limit of the Taylor coefficients. 
From a small number of low-order coefficients, the ratios that one can form are neither a lower nor an upper bound of any
sort on the radius of convergence. In fact, \cite{Gavai:2008zr} considers the Taylor expansion of $\chi_q$ rather than that of the pressure itself. This leads to an equivalent expression for the radius of convergence
\bea
\frac{\chi_q}{T^2} = \frac{1}{T^2} \frac{d^2 P}{d\mu^2} =  \displaystyle\sum_{n=1}^\infty { 2n (2n-1)}~ c_{2n}(T) \left( \frac{\mu}{T} \right)^{2n-2} \\
\frac{\mu_E}{T_E} = \displaystyle\lim_{n\to\infty} \sqrt{\left|\frac{2n (2n-1) c_{2n}(T)}{(2n+2) (2n+1) c_{2n+2}(T)}\right|}
~~~ {\rm at} ~~~ T=T_E
\label{eq:radius_GG}
\eea
But for $n=1$, the ``effective radius of convergence'' differs from that obtained from eq.~(\ref{eq:radius}) by a factor 
$\sqrt{6}$. Thus, the coincidence of estimates from successive small values of $n$ depends in part on the choice of observable
to expand.
Moreover, the QCD critical point is in the universality class of the 3d Ising model, with known, non-trivial critical 
exponents, and the susceptibility $\chi_q$ does not simply diverge like $(\mu_E^2 - \mu^2)^{-1}$, leading to a modification
of eq.~(\ref{eq:radius_GG}) for finite $n$.

Another difficulty is that the Taylor coefficients vary with the temperature, and so does the radius of convergence of the 
expansion. At high temperature, we know from Roberge and Weiss~\cite{Roberge:1986mm} that there is a first-order transition line
at $\mu/T = i \pi/3$. This singularity at negative $\mu^2$ is consistent with the measured Taylor coefficients,
which alternate in sign at high temperature starting from $c_4$. 
At low temperature, a first-order transition occurs at real $\mu$, which should cause high-order Taylor coefficients
to all be positive. This remains true as $T$ increases, until the critical endpoint of the first-order line is 
reached at $T=T_E$.
When $T$ rises above $T_E$, the real singularity due to the critical point branches into
a pair of conjugate poles in the complex $\mu$ plane, as shown by Stephanov in \cite{Stephanov:2006dn}. These complex poles move
towards the imaginary $\mu$ axis as $T$ increases, and approach the origin closer than the QCD critical point~\cite{Stephanov:2006dn}.
Therefore, the determination of $T_E$ should not be based on the minimization of the radius of convergence, but on the
sign behaviour of the Taylor coefficients.
The two groups use different prescriptions. Karsch et al. determine $T_E$ as the highest temperature at which all measured
Taylor coefficients are positive, as appropriate for a singularity at real $\mu$. Gavai and Gupta choose for $T_E$ the {\em lowest}
temperature $T<T_c$ for which all measured Taylor coefficients are positive (see Fig.~11 of \cite{Gavai:2008zr}).

Finally, the strongest reason for skepticism, in my opinion, comes from the original reweighting study of Fodor and Katz~\cite{Fodor:2004nz}. The observable they focused on, the imaginary part of the Lee-Yang zero closest to the real axis, extrapolated
to the thermodynamic limit, is closely related to the inverse of the maximum value of the specific heat. 
In the small-$\mu$ region where reweighting can be trusted, the specific heat shows complete
insensitivity to $\mu$. It would take a high-order Taylor expansion about $\mu=0$ of the curve Fig.~\ref{fig:FK} {\em right} to capture the zero signaling the QCD critical point.
Note, as further confirmation of this lack of sensitivity to a putative critical point, that all susceptibilities measured in the determination of the pseudo-critical line $T_c(\mu)$~\cite{Endrodi} reported Sec.~\ref{sec:Tc} appear to slightly {\em decrease} as $\mu$ is turned on.

\subsection{The curvature of the critical surface}

Considering the difficulties and ambiguities of determining the convergence radius of the Taylor expansion, it is worth
pursuing a complementary, broader strategy: the determination of the chiral critical surface Fig.~\ref{fig:critpt} {\em right},
following the second arrow.
I have been pursuing this approach for several years with Owe 
Philipsen~\cite{de Forcrand:2002ci,de Forcrand:2003hx,deForcrand:2006pv,deForcrand:2008vr,Moscicki:2009id}. 
We have now reached conclusive results on coarse $N_t=4$ lattices, using standard staggered fermions and setting aside possible issues with taking fractional powers of the fermion determinant.

\begin{figure}
\centerline{
\includegraphics[width=5.7cm]{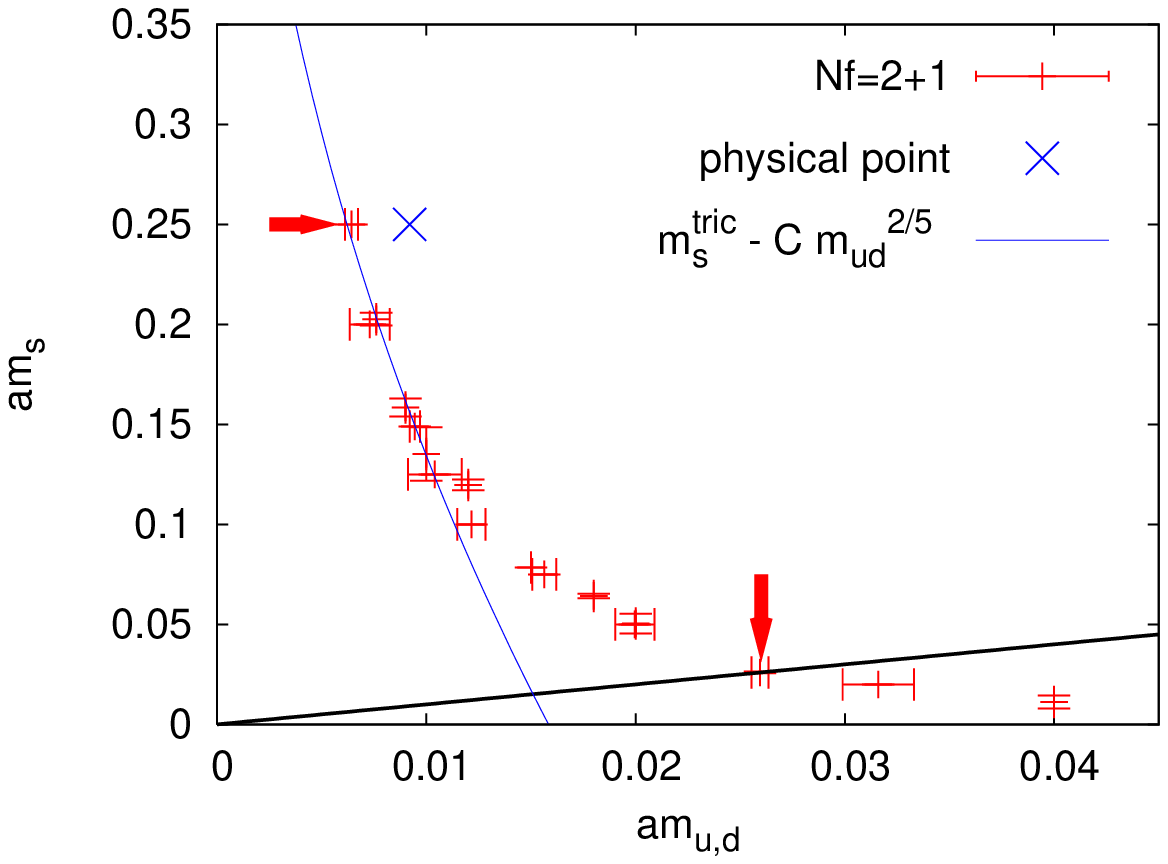}
\put(-30,29){\tiny $N_f=3$}
\hspace*{-0.3cm}
\includegraphics[width=0.40\textwidth]{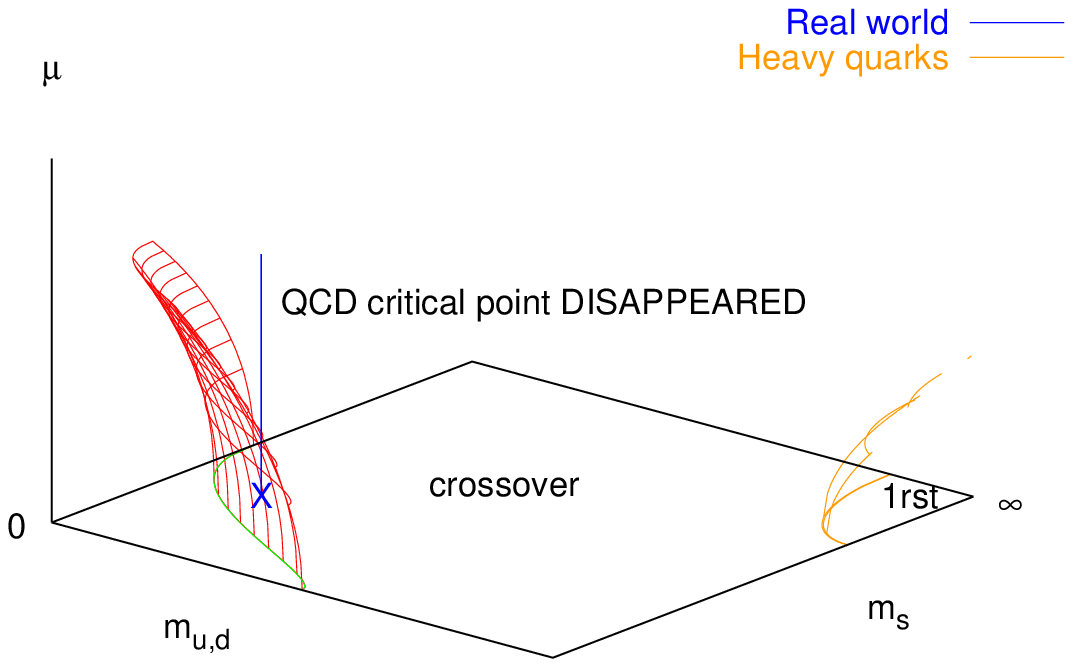}
\hspace*{-0.7cm}
\includegraphics[width=0.40\textwidth]{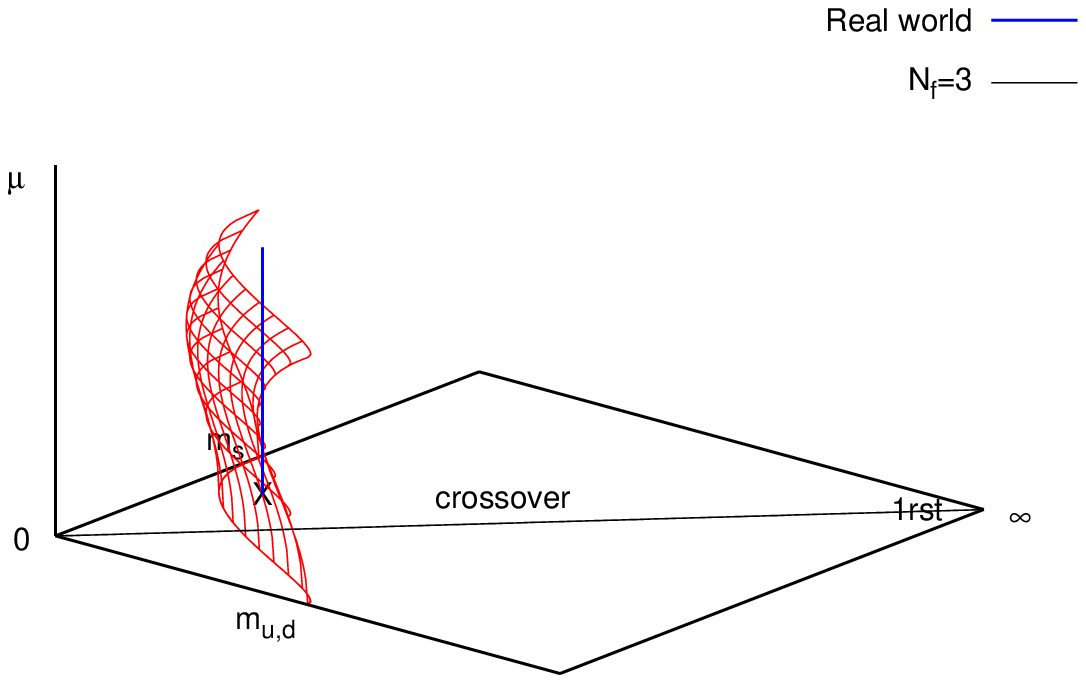}
}
\caption{{\em Left}: Chiral critical line at $\mu=0$ in the $(m_{u,d},m_s)$ quark mass plane~\cite{deForcrand:2006pv}. Compare with expectations Fig.~\protect\ref{fig:critpt} {\em left}. The two red arrows mark the points at which the curvature of the critical surface was measured. 
{\em Center}: Leading effect of a negative curvature. Note the curvature of the deconfinement critical surface for heavy quarks~\cite{Kim:2005ck}. {\em Right}: Possible back-bending due to higher-order terms~\cite{Chen:2009gv}.}
\label{fig:critsurf}
\end{figure}

The first step was the determination of the $\mu=0$ critical line in the $(m_{u,d},m_s)$ quark mass plane, shown Fig.~\ref{fig:critsurf} {\em left}. No surprises were encountered there. We confirmed that the physical point lies in the
region where a finite-temperature crossover takes place, not a phase transition. And our results were consistent with the
tricritical scaling implied by a tricritical point at $m_{u,d}=0$, $m_s \sim 500$ MeV (solid blue curve in Fig.~\ref{fig:critsurf} {\em left}).

The next step has been to measure the variation of the critical quark masses with chemical potential, expressed in Taylor series form as
\be
\frac{m_c(\mu)}{m_c(0)} = 1 + \sum_{k=1} {\bf c_k} \left(\frac{\mu}{\pi T}\right)^{2k}
\ee
This endeavour has been pursued at the two mass points marked by red arrows Fig.~\ref{fig:critsurf} {\em left}, corresponding to three degenerate flavors and to a strange quark with physical mass, respectively. In the $N_f=3$ case, we have compared the
effect of an infinitesimal imaginary $\mu$, which yields the Taylor coefficients $c_k$ directly, and that of several finite imaginary $\mu$'s with results fitted by a polynomial, finding good consistency. We have also compared two spatial volumes, $8^3$ and $12^3$, finding excellent agreement.
Our final result~\cite{deForcrand:2008vr} is
\be
\frac{m_c(\mu)}{m_c(0)} = 1 { -3.3(3)} \left(\frac{\mu}{\pi T}\right)^2 \!\!\! 
{ -47(20)} \left(\frac{\mu}{\pi T}\right)^4  \! {\bf -} \dots 
\label{eq:N_f=3}
\ee
Since we identify the critical mass $m_c$ as that which gives the Binder cumulant $B_4\equiv \frac{\langle (\delta\bar\psi\psi)^4 \rangle}{\langle (\delta\bar\psi\psi)^2 \rangle^2}$ of the quark condensate $\bar\psi\psi$ its critical Ising value 1.604.., our observable is built from $4^{th}$ derivatives of the pressure. Extracting $c_2$ as above then requires
the measurement of $8^{th}$ derivatives of the pressure. As indicated Sec.~\ref{sec:Taylor}, this is the current state of the art. To obtain such results, we accumulated about 25 million RHMC trajectories, and started to use the EGEE computing Grid~\cite{deForcrand:2008zi}.

In the $N_f=2+1$ non-degenerate case with physical strange quark mass, we had to tune $m_{u,d}$ to values smaller than in nature in order to turn the finite-temperature crossover into a second-order phase transition. This forced us to use a larger, $16^3$, volume and use the computing Grid extensively. Our final result~\cite{Moscicki:2009id}, based on about 1.5 million trajectories, is
\be
\frac{m^{u,d}_c(\mu)}{m^{u,d}_c(0)} = 1 { -39(8)} \left(\frac{\mu}{\pi T}\right)^2 \!\!\! {\bf -} \dots
\label{eq:N_f=2+1}
\ee
Thus, in both cases we find that the region of quark masses where a first-order transition takes place {\em shrinks} as
$\mu$ is turned on, just like in the case of heavy quarks where the sign problem is mild and the critical surface can be determined by brute force reweighting~\cite{Kim:2005ck,Langelage:2009jb}. The chiral critical surface then bends as shown Fig.~\ref{fig:critsurf} {\em center}. As indicated in the figure, this rules out, on $N_t=4$ lattices, the presence of a chiral critical point, at least for $\mu/T \lsim {\cal O}(1)$ where one would expect the truncation error of the Taylor expansion to be small.

This conclusion must be accompanied by important cautionary remarks.
First, one may be worried about the convergence of the Taylor expansion. Already for $\mu/T=1$, the last terms of eqs.~(\ref{eq:N_f=3}) and (\ref{eq:N_f=2+1}) are dominant. Even though the sign of the next-order contribution has been estimated
and reinforces the shrinking of the first-order region,
it is clear that higher-order terms may quickly produce a ``back-bending'' of the critical surface as in Fig.~\ref{fig:critsurf} {\em right}. Indeed, this must happen if one wishes to reconcile our result with that of Fodor and Katz~\cite{Fodor:2004nz}, whose critical point Fig.~\ref{fig:FK} lies at $\mu/T \sim 0.7$ for mass parameters similar to eq.~(\ref{eq:N_f=2+1}).
Similarly, there is no inconsistency between our conclusion and that of Ejiri~\cite{Ejiri:2008xt}, who finds a critical point
at $\mu/T \sim 2.4$.
Such a back-bending can also be explained by model calculations. In the $N_f=2$ case,
it is known that a restoration of the $U(1)_A$ symmetry favors a first-order transition~\cite{Pisarski:1983ms,Chandrasekharan:2007up}.
Similarly for $N_f=2+1$, a back-bending surface can be produced in an NJL-type model, by adding a 't Hooft $U(1)_A$-breaking term $(\mathrm{det}\,\bar{q}_{i}(1-\gamma _{5})q_{j}+\mathrm{h.c.})$, with a $\mu$-dependent strength~\cite{Chen:2009gv}~\footnote{Interestingly, back-bending does not necessarily produce a critical point. The critical temperature
is determined implicitly as one moves on the critical surface, and may decrease to zero, thus terminating the critical surface,
before the quark masses have reached their physical values~\cite{Chen:2009gv}.}.
A similar pattern can even be obtained in a linear sigma model including thermal fluctuations~\cite{Bowman:2008kc}.

\begin{figure}
\centerline{
\includegraphics[width=0.336\textwidth]{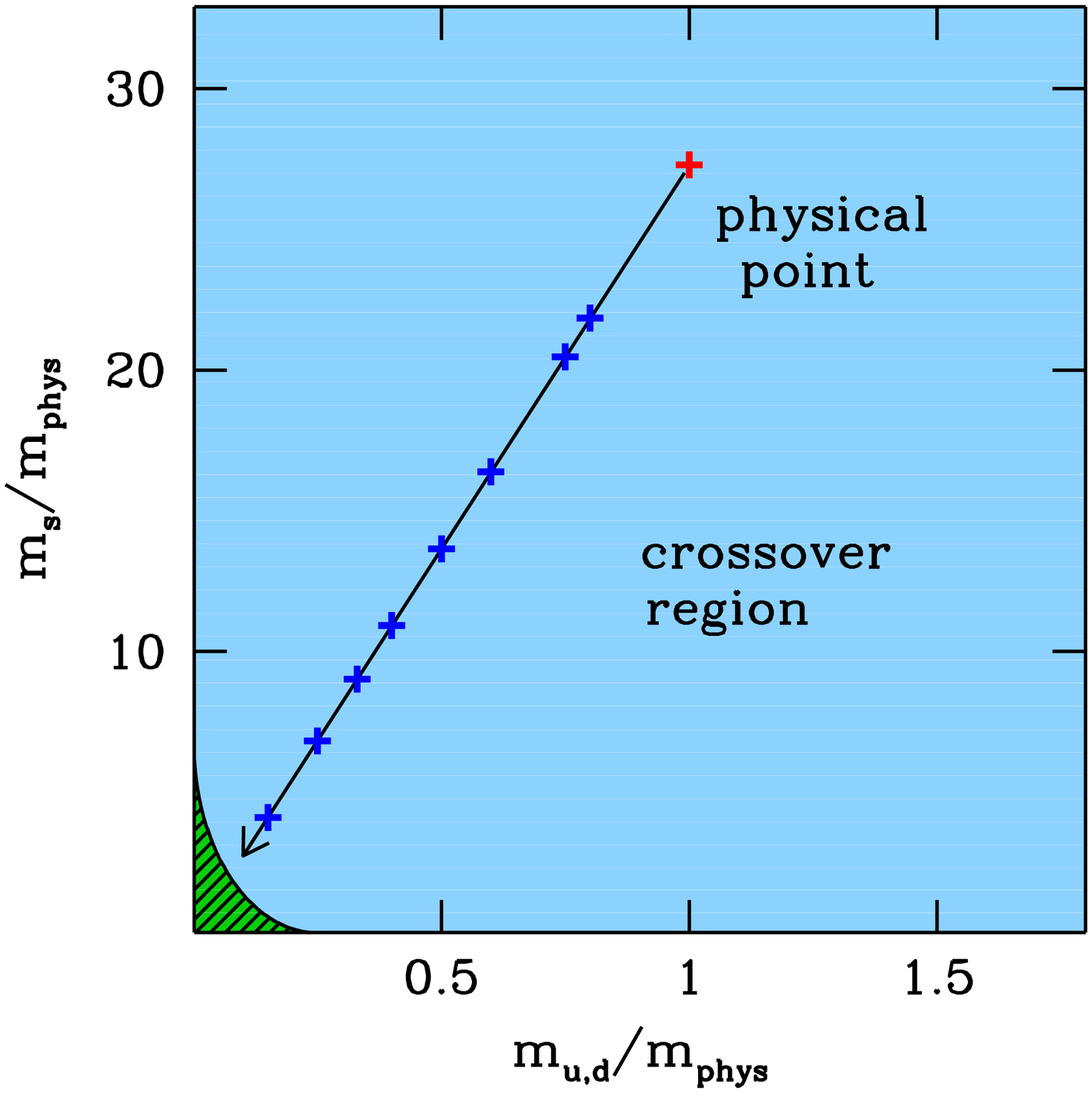}
\put(-127,12){\color{red} \tiny first-order}
\hspace*{1.0cm}
\includegraphics[width=0.50\textwidth]{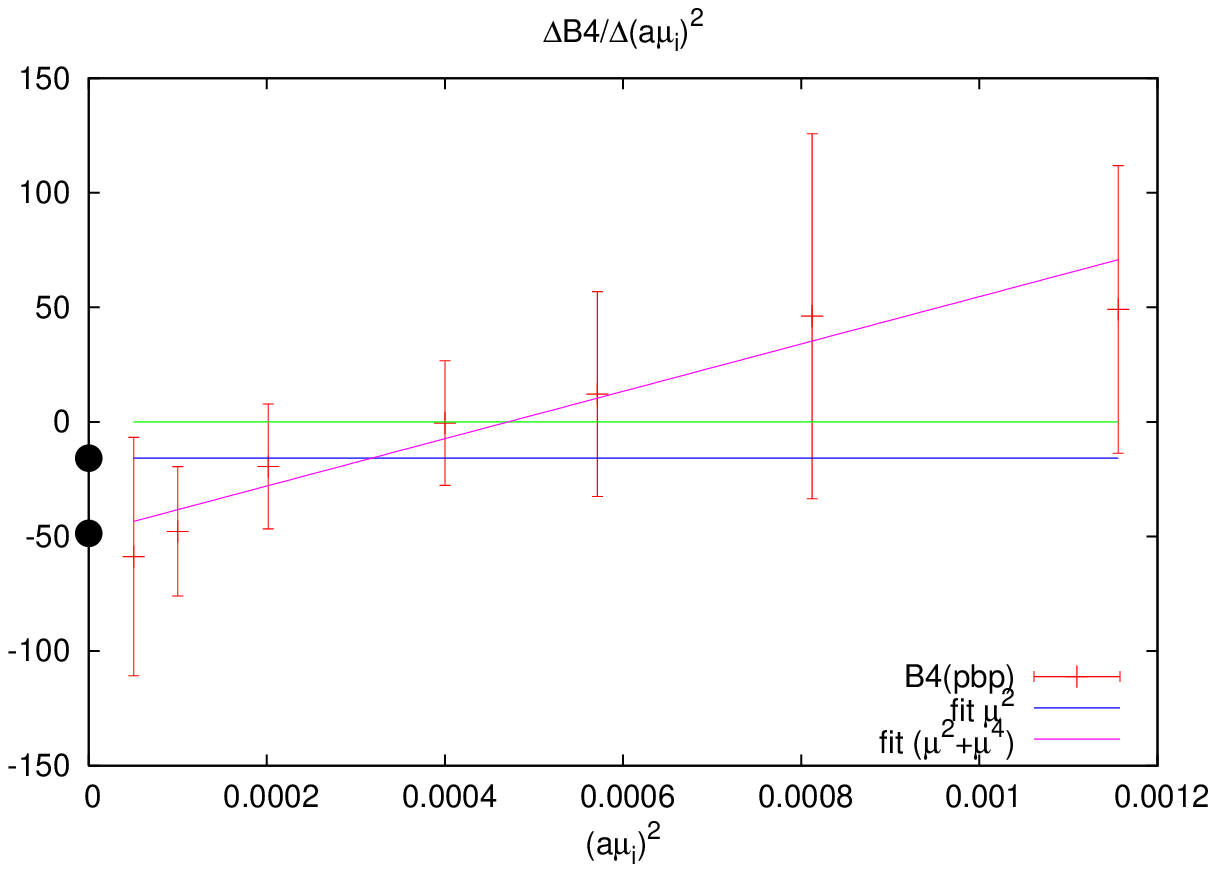}
\put(-145,45){\small \color{red} current status $N_f=3, N_t=6$}
}
\caption{{\em Left}: estimate of the region of first-order transition in the quark mass plane, in physical units,
from improved (stout-smeared) actions on $N_t=4$ and $6$ lattices~\cite{Endrodi:2007gc}. {\em Right}: current status
of the measurement of the curvature of the critical surface on $N_t=6$ lattices. The ${\cal O}((\mu/T)^2)$ Taylor 
coefficient is obtained from the intercept, and the ${\cal O}((\mu/T)^4)$ coefficient from the slope of the fit.
A negative curvature is favored as for $N_t=4$, but the results are not conclusive yet.}
\label{fig:Nt6}
\end{figure}

The second issue is the systematic error caused by the rather coarse lattice spacing: $N_t=4$ implies
$a \sim 0.3$ fm. Even at $\mu=0$ and on $N_t=8$ lattices, discretization errors are presumably the cause of the 
${\cal O}(15\%)$ discrepancy between estimates of
$T_c$ by Karsch et al.~\cite{Bazavov:2009zn} and by Fodor et al.~\cite{Fodor_Tc}.
Here, the critical surface which we study is sensitive to deviations from the Stefan-Boltzmann law at high temperature,
and even more so to the violation of taste symmetry which strongly affects the thermodynamics of the 16 ``pions''.
It should come as no surprise -- a posteriori -- that the critical line in the
quark mass plane, using physical coordinates like $m_\pi/T_c$, 
seems to move by ${\cal O}(100\%)$ as $N_t$ increases from 4 to $\infty$.
Early indications came from improving the discretization of the Dirac operator at fixed $N_t=4$: 
the critical pion mass seemed to decrease by a factor $\sim 4$~\cite{Karsch:2003va}.
More recently, an even more dramatic shrinking of the first-order region 
has been reported in \cite{Endrodi:2007gc}, estimating from stout-smeared $N_t=4$ and $6$ simulations that the
first-order region is limited to quark masses about 10 times smaller than physical (see Fig.~\ref{fig:Nt6} {\em left}).
A less dramatic, but similar effect has been seen in \cite{deForcrand:2007rq} when comparing $m_\pi/T_c$ for the
3-flavor theory on lattices with $N_t=4$ ($m_\pi/T_c = 1.680(4)$) and 6 ($m_\pi/T_c = 0.954(12)$) time-slices.
Clearly, finer lattices tend to make the finite-temperature transition smoother.
This can in fact be seen in an NJL model by truncating the sum over Matsubara frequencies to the first $N_t$ values~\cite{Chen:2009mv}.
This effect by itself makes it more difficult to imagine a critical point at small $\mu$: the physical
point lies further from the critical line, and the critical surface would have to bend very strongly
towards larger quark masses.

Moreover,
one may hope, because the chemical potential does not affect the UV physics, that while the critical
surface will move significantly towards the origin $m_{u,d}=m_s=0$ as $a\to 0$, its {\em curvature}
will vary less. This pious wish needs to be confirmed by numerical simulations, which have proved to be challenging.
Using the same method as for $N_t=4$, we have been simulating an $18^3\times 6$ lattice, with 3 degenerate
quark flavors at the $\mu=0$ critical mass. After over two years of simulation and a half-million units of Molecular Dynamics 
time, the current results
Fig.~\ref{fig:Nt6} {\em right} are still too noisy to conclude. The curvature of the critical surface is obtained from
the $\mu=0$ intercept of the data: a negative value is favored, as for $N_t=4$, but 
the errors are large and the (correlated) data do not exclude a positive value. Moreover, a linear fit, 
including a $(\mu/T)^4$ term, is preferred over a constant fit. The sign of the $(\mu/T)^4$ coefficient then
reinforces the shrinking of the first-order region as in eq.~(\ref{eq:N_f=3}).
However, its magnitude (which is very poorly determined at the moment) could
make this term dominant over the leading $(\mu/T)^2$ term as soon as $\mu/T \gsim 0.15$. 
If this turns out to be the case, then our truncated Taylor expansion can only be trusted up to such values.
Going to larger $\mu$ would require determining higher-order Taylor coefficients, a daunting task.


\section{Left out}

I have not covered several recent developments in the numerical study of finite-density QCD. \\
$\bullet$ The numerical study of the canonical ensemble, with a fixed number of baryons, has been extended from staggered
fermions~\cite{Hasenfratz:1991ax,deForcrand:2006ec,Ejiri:2008xt} to Wilson fermions, with several technical refinements~\cite{Alexandru:2007bb,Meng:2008hj,Danzer:2008xs}.
This is computationally challenging, since the fermion determinant must be computed exactly, at a cost proportional to
the cube of the matrix size, so that the work is increased 64-fold on a given lattice size.
On a $6^3\times 4$ lattice, the phase diagram appears to be qualitatively different for $N_f=2, 3$ and $4$ flavors, suggesting a possible critical point for $N_f=3$. These results are presented by Anyi Li in a plenary talk~\cite{Li:2010dy}. \\
$\bullet$ Instead of aiming at a finite density of baryons, one may study $T=0$ few-body physics: by measuring the interactions between
two or three baryons, one can constrain the couplings of an effective theory describing nuclear matter. This is a very important
and active research direction, reviewed last year~\cite{Beane:2008ia}, with two large-scale efforts underway~\cite{Aoki:2009ji,Beane:2009py}. The sign problem appears in the form
of a signal-to-noise ratio for baryon correlators degrading exponentially fast in Euclidean time~\cite{Lepage}. But since
one works at $T=0$, a toolbox of variational trial states can be used to try and isolate the groundstate before the signal
dies away~\cite{Beane:2009gs}. The state of the art is reviewed by Will Detmold in a plenary talk~\cite{Detmold}. Spectacular results have already been obtained for multi-meson states, corresponding to an isospin chemical potential~\cite{Beane:2007es}. \\
$\bullet$ Other interesting developments concern non-QCD theories, e.g. 2-color ``QCD''~\cite{Hands:2010gd} where a variety of regimes may appear at low temperature, as the chemical potential is increased, or $(1+1)d$ Gross-Neveu and NJL
models whose phase diagram can be determined analytically~\cite{Basar:2008ki,Basar:2009fg}, including inhomogeneous, crystalline phases. These features may well be present in QCD also. \\
$\bullet$ An important approach to tackling the sign problem is the ``density of states'' approach, where the partition function
is expressed as a one-dimensional integral $Z = \int dx~\rho(x)$, and the computer effort can be concentrated on values of $x$
where $\rho(x)$ is more noisy~\cite{Bhanot:1986kv}. A large-scale effort, taking for $x$ the gluon action, gave hints of a triple point
in the $(\mu,T)$ plane~\cite{Fodor:2007vv}. More recently, this approach has been espoused, in part, by Ejiri in \cite{Ejiri:2008xt}. \\
$\bullet$ Several contributions to this Conference did not fit in the categories above, and I omitted them from this subjective review,
with apologies.
    

\section{Prospects}

I have painted a rather bleak landscape, where systematic errors are very large and one should only expect slow progress, 
helped with massive amounts of computer time. Is there nothing more exciting in sight?
Yes, definitely. Let me emphasize two directions where I consider enthusiasm to be justified.
Amusingly, they both represent a revival of topics which were ``hot'' twenty years ago.

\subsection{Worldline formalism and strong coupling limit}

As we have seen Sec.~\ref{sec:sign}, as soon as one integrates out the fermions, one must encounter a sign problem with
the resulting determinant at finite density. This suggests changing the order of integration, and performing at least
a partial integration over the gauge fields first~\cite{Bringoltz:2010iy}. Integrating out the gauge links seems hopeless, since the plaquette term in the action introduces a complicated 4-link interaction. Still, this may be simpler than dealing with the sign problem. To start with, one can consider the strong-coupling limit $\beta_{\rm gauge}=0$, where the plaquette term in the action
drops out. Then, the link integration factorizes into a product of 1-link integrals, which can be performed analytically.
Only color singlets survive, made of quark fields. The Grassmann integration produces only a handful of terms, which
represent the hopping of color singlets from site to site. In other words, the partition function has been reexpressed
as a sum over configurations of {\em loops}, representing the worldlines of hadronic color singlets.

At this stage, this reformulation is relatively simple, since it only involves discrete variables, and physically appealing.
But it is not clear that simulations will be any easier: The fixed number of underlying Grassmann variables at each site
generates a constraint on the loop configuration (the number of $\bar{q}q$ mesons connected to each site is fixed), which
makes local Monte Carlo updates hopelessly inefficient. And a severe sign problem follows from the fermionic nature of the
baryons (for an odd number of colors): baryon loops are oriented, and their weight flips sign with their orientation.
Fortunately, this sign problem can be solved by partial resummation at $\mu=0$, and then remains very mild at $\mu\neq 0$~\cite{Karsch:1988zx}.
And a recent Monte Carlo algorithm, the ``worm'' algorithm~\cite{Prokof'ev:2001zz}, is particularly efficient at updating globally
such discrete systems with constraints~\cite{Adams:2003cca}. This efficiency does not even degrade in the chiral limit.

All this was understood and tried around 1990~\cite{Karsch:1988zx,Boyd:1991fb}, and abandoned before the discovery of the worm algorithm.
This crucial algorithmic progress has enabled a complete numerical determination of the $(\mu,T)$ phase diagram for the strong coupling limit of QCD with staggered fermions~\cite{deForcrand:2009dh}, superseding ancient~\cite{Damgaard:1985bn} and recent~\cite{Miura:2008gd}
analytical approximations, and even a precise study of nuclear interactions and of ``nuclear matter'' in this lattice model,
all with ``tabletop'' computer resources.

I am insisting on one particular project because of my personal involvement. But the same worldline approach can be applied to other lattice fermion models~\cite{Gattringer:2007em,Wenger:2009mi}, and to bosonic models normally afflicted by a sign problem, after transforming to dual variables~\cite{Banerjee:2010kc}. It may even be useful to ``fermionize'' a bosonic system, treating the bosons as fermion
composites. Recall that this yields a simple derivation of Onsager's solution to the $2d$ Ising model~\cite{Samuel:1978zx}. 
This ``new computational approach'' was reviewed by its main proponent at last year's Conference~\cite{Chandrasekharan:2008gp}.

Can all lattice models, in particular QCD at weak coupling, be formulated and efficiently simulated as a gas of loops?
One technical ingredient for success is the worm algorithm. Its limits of applicability are not clear yet. It appears to work well for at least one model, the $2d$ $CP^{N-1}$ spin model~\cite{Wolff:2010qz}, where cluster algorithms are known to fail.
If the random worms can be generalized to random surfaces, then one could apply the same treatment to the Yang-Mills part of
the action and simulate QCD at weak coupling, as a gas of quark loops forming the boundary of gauge surfaces.
Unfortunately, this step requires a duality transformation, which for a non-Abelian theory gives negative weights and/or
non-local couplings~\cite{Ukawa:1979yv,Cherrington:2007ax}.

A somewhat less ambitious strategy consists of designing fermion-based, sign-problem free actions, with symmetries which ensure  a desired effective low energy limit. Such actions can be efficiently simulated, even in the massless limit, to address 
precise questions about the effective low energy theory. A variety of models with chiral symmetry can be realized~\cite{Cecile:2007dv}. What is still missing is a non-Abelian gauge symmetry.

\subsection{Complex Langevin}

The idea of stochastic quantization is to introduce a Langevin evolution for a field $\phi$ in a fictitious time $\tau$, obeying
\be
\frac{\partial\phi}{\partial\tau} = -\frac{\delta S\left[\phi\right]}{\delta\phi} + \eta
\label{eq:Langevin1}
\ee
where $\eta$ is a Brownian noise. When the action $S\left[\phi\right]$ is real, one can prove the existence of an
associated Fokker-Planck equation and its convergence to the fixed-point distribution $\propto \exp(-S\left[\phi\right])$,
so that all observables satisfy
\be
\langle W\left[\phi\right] \rangle_\tau = \frac{1}{Z} \int {\cal D}\phi \exp(-S\left[\phi\right]) W\left[\phi\right]
\label{eq:Langevin2}
\ee
where $\langle .. \rangle_\tau$ is an average over the fictitious Langevin time $\tau$.

What happens when $S\left[\phi\right]$ is complex? The drift force $-\frac{\delta S\left[\phi\right]}{\delta\phi}$ becomes
complex, so that each component of the field $\phi$ will become complex under the evolution eq.~(\ref{eq:Langevin1}).
One can then complexify $\phi$ into $(\phi^{R} + i \phi^{I})$, evolve with the complexified version
of eq.~(\ref{eq:Langevin1}), and monitor the time-average $\langle W\left[\phi^{R} + i \phi^{I}\right] \rangle_\tau$. There still is an associated Fokker-Planck equation, but almost nothing is known about stationary solutions. Nevertheless,
it turns out that $\langle W\left[\phi^{R} + i \phi^{I}\right] \rangle_\tau$ has the desired value
$\frac{1}{Z} \int {\cal D}\phi \exp(-S\left[\phi\right]) W\left[\phi\right]$ -- sometimes. Some other times, the 
$(\phi^{R} + i \phi^{I})$ distribution in the complex plane does not converge to a $\tau$-stationary
distribution, because $(\phi^{R} + i \phi^{I})$ runs to infinity. And some other times, convergence
is achieved to a distribution giving wrong expectation values $\langle W\left[\phi^{R} + i \phi^{I}\right] \rangle_\tau$. An understanding of sufficient conditions (besides taking a real action) to avoid runaways and convergence to the wrong answer was not reached in the 1980s~\cite{Parisi:1984cs,Klauder:1983zm,Karsch:1985cb,Ambjorn:1985iw}. Activity on this topic died away.

It was noticed only recently that a drastic reduction in the discrete stepsize of the complex Langevin evolution sufficed to
almost eliminate runaways. With an adaptive stepsize~\cite{Aarts:2009dg} the runaways disappeared completely. This opened a new playground where to study convergence. Surprisingly, correct answers have been obtained in many cases: toy models with one gauge link matrix; $4d$ complex $\phi^4$ theory with a chemical potential, where $T\approx 0$ Bose-Einstein condensation is reproduced for $\mu\geq\mu_c$~\cite{Aarts:2008wh}, as well as the lack of $\mu$-induced effect for $\mu < \mu_c$ (known as the ``Silver Blaze problem''~\cite{Cohen:2003kd}); and even QCD with chemical potential, in the heavy-dense limit, where results are consistent with
those of reweighting~\cite{Aarts:2008rr}. These extraordinary successes should be balanced against a short list of failures:
the noise $\eta$ in eq.~(\ref{eq:Langevin1}), which a priori only needs to satisfy $\langle \eta(\tau) | \eta(\tau') \rangle = 2 \delta(\tau-\tau')$ and can be complex, must in fact be kept real for convergence; and
real-time quantum evolution still seems to be out of reach~\cite{Berges:2006xc}.
Clearly, the approach looks promising and its limits must be further explored and understood.

\begin{figure}
\centerline{
\includegraphics[width=0.75\textwidth]{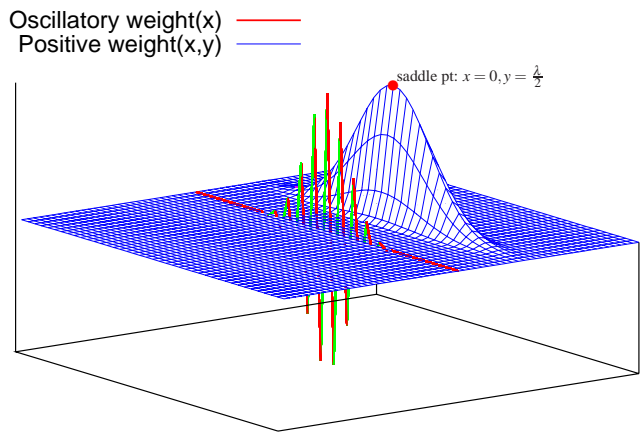}
\put(-139,169){\small\color{red} $\bullet$}
\put(-135,173){\tiny saddle pt: $x=0, y=\frac{\lambda}{2}$}
}
\caption{Complex Langevin for 1 degree of freedom $x \to (x + i y)$: the oscillatory $x$-distribution of Fig.~\protect\ref{fig:2d_oscillatory}, shown in red and green, becomes a smooth positive Gaussian in the $(x,y)$ plane,
centered at the complex saddle point $i \lambda/2$. All moments of the original, oscillatory $x$-distribution are equal
to moments of $(x + i y)$ with respect to the Gaussian, positive $(x,y)$ distribution.}
\label{fig:Langevin}
\end{figure}

To get the flavor of the magic at work here, let us consider the simple example of Sec.~\ref{subsec:sign}:
$Z(\lambda) \equiv \int_{-\infty}^{+\infty} dx \exp(-x^2 + i\lambda x)$. When $\lambda=0$, the real Langevin evolution
is $dx/d\tau = -2 x + \eta$. When $\lambda \neq 0$, the drift force becomes complex and one needs to complexify $x$
into $(x + i y)$. The corresponding Langevin evolution becomes
\be
\frac{d}{d\tau}({x+iy}) = -2({x+iy}) + i\lambda + \eta
\ee
In this simple case, this equation can be solved analytically. Since $Z(\lambda)$ is a Gaussian integral, the stationary distribution of $(x,y)$ is a nice Gaussian shown Fig.~\ref{fig:Langevin}, centered at the complex saddle point $(x=0,y=i\lambda/2)$. 
It is perhaps not intuitive how expectation values $\langle W(x) \rangle$ are recovered: one must analytically continue
$W(x)$ to $W(x+iy)$ to obtain $\langle W(x+iy) \rangle_{\tau} = \langle W(x) \rangle_Z$. The original, negative contributions
of some values of $x$ are then reconstructed when one integrates $W(x+iy)$ over $y$.
Note that the $y$-width of the Gaussian depends on the variance of the imaginary part of the Langevin noise: for a real noise,
the $y$-width shrinks to zero. But, in this case at least, correct answers are obtained for any complex Langevin noise
satisfying $\langle \eta(\tau) | \eta(\tau') \rangle = 2 \delta(\tau-\tau')$.

This toy example suggests that an analysis of the saddle points of the classical action can be fruitful. This is the starting
point of a loop-like expansion considered in \cite{Guralnik:2009pk}, which may shed light on the convergence properties of complex Langevin: The Langevin noise distorts the classical Gaussian distribution around the saddle points, hopefully in a weak manner.
Indeed, one may guess that systems with one complex saddle point can perhaps be safely studied by complex Langevin, 
while competition between saddle points (as at a first-order transition) may present a challenge. The ``safe'' category might
include the effect of a $\theta$ vacuum angle, as studied in a saddle-point formulation in \cite{Azcoiti:2002vk}.
Gauge theories are more likely to be in the ``unsafe'' category, due to flat directions in the action, which correspond
to gauge transformations and extend to complex infinity because gauge links are complexified from $SU(N)$ to $SL(N,{\bf C})$.
Progress is being made towards understanding necessary conditions for convergence~\cite{Aarts:2009uq}.
Finally, I note the intriguing, perhaps fruitful analogy between complex Langevin and $PT$-symmetric quantum mechanics~\cite{Bender:2005tb} and its cousin, complexified classical mechanics~\cite{Bender:2007pr}.


\section{Conclusions}

Determining the phase diagram of QCD as a function of temperature and chemical potential is an important fundamental goal.
Even if clear answers are not available yet, and if the progress in our knowledge is likely to be slow,
it is definitely worth pursuing the present efforts. Finite-density lattice QCD is not just a temporarily fashionable topic:
it justifies a sustained research program.

Three numerical approaches give reliable, consistent results provided the chemical potential is small enough:
reweighting, Taylor expansion and analytic continuation from imaginary $\mu$. This allows crosschecks in the
region where the reliability of these methods becomes doubtful. Confidence in finite-density simulations can be
established, so that comparison between QCD and models can become reliable and fruitful.
For instance, the well-established phase diagram of QCD at imaginary $\mu$ is already a useful testing ground for
effective models.

Similarly, the curvature of the pseudo-critical temperature $T_c(\mu)$ at $\mu=0$ is almost under numerical control,
and provides useful phenomenological information.
The situation is more delicate regarding the existence and location of a QCD critical point, which requires venturing
to non-zero $\mu$. Nevertheless, analytic knowledge about the severity of the sign problem can give us, before we start
the computation, reliable information about the $(\mu,T)$ region which can be explored and the baryon density which can
be reached. And it now seems clear that the coarseness of the usual $N_t=4$ or $6$ lattices is the major source of
systematic error. Thus, with necessary massive increases in computer resources, we can expect slow but steady progress towards the final, continuum limit answer.

This review suggests two possibilities for breakthroughs, or at least for rapid development: reversing the order of integration and integrating over the gauge links first; and dealing with the sign problem using complex Langevin. Whatever the final
scope of these two strategies, one can already predict that they will contribute to increasing our knowledge of finite-density
QCD at least in some limiting regimes of parameters.

I would like to end by citing Confucius, who knew all about the importance of scholarly research:
``Real knowledge is to know the extent of one's ignorance''.

\section*{Acknowledgements}

I am very grateful to numerous colleagues for discussions and patient explanations, among them
G.~Aarts,
A.~Alexandru,
M.~Alford,
B.~Bringoltz,
M. D'Elia,
W.~Detmold,
S.~Ejiri,
Z.~Fodor,
M.~Fromm,
K.~Fukushima,
C.~Gattringer,
R.~Gavai,
M.~Goltermann,
S.~Gupta,
S.~Hands,
A.~Hasenfratz,
S.~Kim,
A.~Kurkela,
K.-F.~Liu,
A.~Ohnishi,
M.~Panero,
O.~Philipsen,
C.~Schmidt,
S.~Sharpe,
K.~Splittorff.


\end{document}